\documentclass[aps,prd,twocolumn,superscriptaddress]{revtex4}
\usepackage{epsfig,epsf}
\usepackage{amsmath}
\usepackage{amsthm}
\usepackage{amsfonts}
\usepackage{amssymb}
\usepackage{dsfont}
\usepackage{multirow}
\usepackage{appendix}
\usepackage{slashed}
\usepackage[active]{srcltx}
\usepackage{psfrag}

\begin{document}

\title{Heavy scalar molecule $B_{c}^{+} B_{c}^{-}$}
\date{\today}
\author{S.~S.~Agaev}
\affiliation{Institute for Physical Problems, Baku State University, Az--1148 Baku,
Azerbaijan}
\author{K.~Azizi}
\thanks{Corresponding Author}
\affiliation{Department of Physics, University of Tehran, North Karegar Avenue, Tehran
14395-547, Iran}
\affiliation{Department of Physics, Dogus University, Dudullu-\"{U}mraniye, 34775
Istanbul, T\"{u}rkiye}
\author{H.~Sundu}
\affiliation{Department of Physics Engineering, Istanbul Medeniyet University, 34700
Istanbul, T\"{u}rkiye}

\begin{abstract}
Mass and full width of the heavy scalar molecule $\mathcal{M}$ composed of
the mesons $B_{c}^{+}$ and $B_{c}^{-}$ are calculated in the QCD sum rule
framework. To find its mass and current coupling, we apply the two-point sum
rule method. The width of the hadronic molecule $\mathcal{M}
=B_{c}^{+}B_{c}^{-}$ is evaluated by taking into account its dissociation to
$\eta_c\eta_{b}, J/\psi \Upsilon$, and $B_{c}^{+}B_{c}^{-} $ mesons. Decays
into $D$ and $B$ meson pairs with appropriate charges and quantum numbers
triggered by $\overline{b}b$ and $\overline{c}c$ annihilations to light
quark-antiquarks are also included in the analyses. Because the partial widths
of $\mathcal{M}$ molecule's decay channels depend on the strong couplings at
$\mathcal{M}$-meson-meson vertices, we estimate them by invoking tools of
the three-point sum rule method. Our predictions $m=(12725 \pm 85)~\mathrm{%
MeV}$ and $\Gamma[\mathcal{\ M}]=(155 \pm 23)~ \mathrm{MeV}$ for the
parameters of the molecule $\mathcal{\ M}$ provide useful information for
experimental studies of numerous heavy resonances.
\end{abstract}

\maketitle


\section{Introduction}

\label{sec:Intro}
Internal organization of multiquark hadrons, their parameters, production
and decay mechanisms are among important questions of the quark-parton model
and quantum chromodynamic (QCD). Two primary motivations drive the
investigation of such structures. Firstly, there is a fundamental
theoretical interest in exploring particles that could exist within the
quark-gluon model framework. Secondly, and more significantly, these yet
hypothetical structures offer a mean to elucidate properties of the
particles and reconcile them with experimental data. For instance, this
concept was utilized in Ref.\ \cite{Jaffe:1976ig} to analyze a mass
hierarchy within the lowest scalar multiplet. According to this work, the
nonet of the light scalars are four-quark states $q^{2}\overline{q}^{2}$
which accounts for the experimentally observed features of these mesons.

Production of numerous $J^{\mathrm{PC}}=1^{--}$ states by $e^{+}e^{-}$
annihilation in the energy range $3.9-4.4~\mathrm{GeV}$ triggered attempts
to explain this phenomenon by existence of exotic $c\overline{q}\overline{c}%
q $ mesons \cite{Bander:1975fb}. These four-quark particles are made of
color-singlet $c\overline{q}$ and $\overline{c}q$ components. Similar
concepts regarding the formation of charmonium molecules were also advanced
by other researchers \cite{Voloshin:1976ap,DeRujula:1976zlg}. In this
scenario, the four-quark mesons manifest as bound or/and resonant states of
the $D$ meson pairs, interacting via conventional light meson exchange.

In subsequent years, the concept of hadronic molecules constructed from
ordinary mesons was further elaborated and advanced in numerous publications
\cite{Tornqvist:1991lks,Ding:2008mp,Zhang:2009vs,Albuquerque:2012rq,
Chen:2015ata,Karliner:2015ina,Liu:2016kqx,Sun:2018zqs,Chen:2017vai,
PavonValderrama:2019ixb,Molina:2020hde,
Xu:2020evn,Xin:2021wcr,Agaev:2022duz,Agaev:2023eyk,Wang:2023bek,Braaten:2023vgs,Wu:2023rrp, Liang:2023jxh,Liu:2023gla,Liu:2024pio,Wang:2025zss,Braaten:2024tbm,Yalikun:2025ssz}, and it still remains an active area of research. These studies explored
the binding mechanisms of such states, calculated their masses using diverse
models and approaches, analyzed processes through which these particles
might be observed.

Over the past two decades, experimental investigations collected significant
and rich information regarding potential candidates to exotic tetraquark
mesons. All such resonances were examined within the frameworks of both
diquark-antidiquark and hadronic molecule models. Prominent among recent
experimental achievements is the discovery of \ four $X$ resonances in di-$%
J/\psi $ and $J/\psi \psi ^{\prime }$ mass distributions by the LHCb, ATLAS,
and CMS collaborations \cite%
{LHCb:2020bwg,Bouhova-Thacker:2022vnt,CMS:2023owd}. These resonances are
located in the mass interval $6.2-7.3~\mathrm{GeV}$ and are presumed to be
hidden-charm scalar tetraquarks $cc\overline{c}\overline{c}$. These fully
charmed four-quark resonances were also investigated in our previous
articles \cite{Agaev:2023wua,Agaev:2023ruu,Agaev:2023gaq,Agaev:2023rpj},
where we employed the QCD sum rule (SR) method to evaluate their masses and
full decay widths. A comparison of our predictions with available
experimental data led to the interpretation of the lightest state, $X(6200)$%
, as an $\eta _{c}\eta _{c}$ molecule \cite{Agaev:2023ruu}, while $X(6600)$
is likely a diquark-antidiquark state with axial-vector constituents \cite%
{Agaev:2023wua}. A superposition of a diquark-antdiquark state and a
hadronic $\chi _{c0}\chi _{c0}$ molecule exhibits parameters consistent with
those of the state $X(6900)$ \cite{Agaev:2023ruu,Agaev:2023gaq}. The
heaviest resonance, $X(7300)$, can be considered as an admixture of a $\chi
_{c1}\chi _{c1}$ molecule and the first radial excitation of $X(6600)$ \cite%
{Agaev:2023rpj}.

Another significant discovery by LHCb which has profoundly impacted exotic
hadron physics, is the axial-vector resonance $T_{\mathrm{cc}}^{+}$,
observed in $D^{0}D^{0}\pi ^{+}$ mass distribution \cite%
{Aaij:2021vvq,LHCb:2021auc}. With a $cc\overline{u}\overline{d}$ content, it
represents the first experimentally discovered open-charm tetraquark
candidate. This state has garnered considerable interest within the high
energy physics community and has been explored using various theoretical
approaches (see, Refs. \cite{Agaev:2021vur,Agaev:2022ast} and references
therein). In our previous works \cite{Agaev:2021vur,Agaev:2022ast}, we
evaluated the mass and decay width of this resonance, by modeling it as both
a $[cc][\overline{u}\overline{d}]$ diquark-antidiquark and a $D^{0}D^{\ast
+} $ molecule state. Our results suggest that $T_{\mathrm{cc}}^{+}$
preferentially possesses a diquark-antidiquark structure, although the
computational accuracy does not preclude its interpretation as a hadronic
molecule.

These well-established experimental findings have demonstrated that fully
heavy tetraquarks are accessible in ongoing experiments, thereby encouraged
research efforts dedicated to the thorough analysis of tetraquark mesons
composed exclusively of $c$ and $b$ quarks. Exotic structures with hidden
charm and bottom, such as $bc\overline{b}\overline{c}$ tetraquarks, form an
interesting family of such particles. These states were investigated
intensively in the literature (see, Refs. \cite%
{Agaev:2024wvp,Agaev:2024mng,Agaev:2025} and references therein). In our
studies \cite{Agaev:2024wvp,Agaev:2024mng,Agaev:2025}, we examined the
scalar, axial-vector, and tensor $bc\overline{b}\overline{c}$ particles
within the diquark-antidiquark framework and calculated their masses and
full decay widths.

In this article, we model the scalar $bc\overline{b}\overline{c}$ tetraquark
as a hadronic molecule $\mathcal{M}=B_{c}^{+}B_{c}^{-}$ and compute its
parameters. The mass $m$ and current coupling $\Lambda $ of $\mathcal{M}$
are found using the two-point sum rule method \cite%
{Shifman:1978bx,Shifman:1978by}. As it will be demonstrated, the hadronic
molecule $\mathcal{M}$ is heavier than two-meson thresholds $\eta _{b}\eta
_{c}$, $J/\psi \Upsilon $, and $B_{c}^{+}B_{c}^{-}$; consequently,
dissociations into these meson pairs represent its kinematically allowed
decay channels. Other mechanisms also exist for decays of $\mathcal{M}$ into
ordinary mesons \cite{Becchi:2020mjz, Becchi:2020uvq,Agaev:2023ara}. Indeed,
owing to $b\overline{b}$ and $c\overline{c}$ annihilations into $\overline{q}%
q$ and $\overline{s}s$ pairs, $\mathcal{M}$ readily decays into $D^{(\ast
)+}D^{(\ast )-}$, $D^{(\ast )0}\overline{D}^{(\ast )0}$, and $D_{s}^{(\ast
)+}D_{s}^{(\ast )-}$, as well as $B^{(\ast )+}B^{(\ast )-}$, $B^{(\ast )0}%
\overline{B}^{(\ast )0}$, and $B_{s}^{(\ast )0}\overline{B}_{s}^{(\ast )0}$
meson pairs. Decay widths for all decay channels of the $\mathcal{M}$
molecule are evaluated within the framework of the three-point sum rule
method, which is necessary for determining the strong couplings at the
corresponding $\mathcal{M}$-meson-meson vertices.

This work contains the following sections: In Sec.\ \ref{sec:Mass}, we
determine the mass and current coupling of the scalar molecule $\mathcal{M}$%
. The fall-apart processes $\mathcal{M}\rightarrow \eta _{b}\eta _{c},\
J/\psi \Upsilon $, and $B_{c}^{+}B_{c}^{-}$ are explored in Sec.\ \ref%
{sec:Widths1}. The decays into mesons $D^{(\ast )+}D^{(\ast )-}$, $D^{(\ast
)0}\overline{D}^{(\ast )0}$, and $D^{(\ast )+}D^{(\ast )-}$ are considered
in Sec. \ref{sec:Widths2}. Partial widths of the processes with two
final-state $B$ mesons are calculated in Sec.\ \ref{sec:Widths3}. Here, we
evaluate the full decay width of the molecule $\mathcal{M}$ as well. Our
conclusions are collected in Sec.\ \ref{sec:Conc}.


\section{Spectroscopic parameters of $\mathcal{M}=B_{c}^{+}B_{c}^{-}$}

\label{sec:Mass}

To derive the SRs for the mass $m$ and current coupling $\Lambda $ of the
scalar molecule $\mathcal{M}$, we consider the correlation function%
\begin{equation}
\Pi (p)=i\int d^{4}xe^{ipx}\langle 0|\mathcal{T}\{J(x)J^{\dag
}(0)\}|0\rangle ,  \label{eq:CF1}
\end{equation}%
where $J(x)$ is the interpolating current for the molecule $\mathcal{M}$,
and $\mathcal{T}$ \ is time-ordered product of currents.

In the hadronic molecule picture the interpolating current for $\mathcal{M}$
is
\begin{equation}
J(x)=[\overline{b}_{a}(x)i\gamma _{5}c_{a}(x)][\overline{c}_{b}(x)i\gamma
_{5}b_{b}(x)],  \label{eq:CR1}
\end{equation}%
where $a$, $b$ are the color indices. This current corresponds to the scalar
particle with the quantum numbers $J^{\mathrm{PC}}=0^{++}$.

The sum rules for $m$ and $\Lambda $ can be found by calculating $\Pi (p)$
using the physical parameters of $\mathcal{M}$ which establishes the
phenomenological component of SRs $\Pi ^{\mathrm{Phys}}(p)$. The correlator $%
\Pi ^{\mathrm{Phys}}(p)$ is given by the expression
\begin{equation}
\Pi ^{\mathrm{Phys}}(p)=\frac{\langle 0|J|\mathcal{M}\rangle \langle
\mathcal{M}|J^{\dagger }|0\rangle }{m^{2}-p^{2}}+\cdots .  \label{eq:Phys1}
\end{equation}%
In Eq.\ (\ref{eq:Phys1}) the term presented explicitly is a contribution of
the ground-state particle, whereas effects of higher resonances and
continuum states are shown by ellipses.

To calculate $\Pi ^{\mathrm{Phys}}(p)$, it is useful to write it using the
parameters $m$ and $\Lambda $. For these purposes, we introduce the matrix
element
\begin{equation}
\langle 0|J|\mathcal{M}\rangle =\Lambda .  \label{eq:ME1}
\end{equation}%
After simple manipulations, one gets
\begin{equation}
\Pi ^{\mathrm{Phys}}(p)=\frac{\Lambda ^{2}}{m^{2}-p^{2}}+\cdots .
\label{eq:Phys2}
\end{equation}
A term in $\Pi ^{\mathrm{Phys}}(p)$ with the simple Lorentz structure $%
\Lambda ^{2}/(m^{2}-p^{2})$ is the amplitude $\Pi ^{\mathrm{Phys}}(p^{2})$
necessary for our analysis.

We calculate $\Pi (p)$ with certain accuracy also by utilizing the heavy
quark propagators and operator product expansion ($\mathrm{OPE}$) . The
function $\Pi ^{\mathrm{OPE}}(p)$ obtained after this computation
establishes the QCD side of the SRs. It contains the perturbative and
nonperturbative terms: The latter being proportional to $\langle \alpha
_{s}G^{2}/\pi \rangle $. The reason is that heavy quark propagators do not
contain light quark and mixed quark-gluon condensates. As a result, next
nonperturbative contributions to $\Pi ^{\mathrm{OPE}}(p)$ are generated by
gluon condensates $\langle g_{s}^{3}G^{3}\rangle $ and $\langle \alpha
_{s}G^{2}/\pi \rangle ^{2}$, which are small and neglected in what follows.

The $\Pi ^{\mathrm{OPE}}(p)$ obtained using propagators of the heavy quarks
is
\begin{eqnarray}
&&\Pi ^{\mathrm{OPE}}(p)=i\int d^{4}xe^{ipx}\mathrm{Tr}\left[ \gamma
_{5}S_{c}^{aa^{\prime }}(x)\gamma _{5}S_{b}^{a^{\prime }a}(-x)\right]  \notag
\\
&&\times \mathrm{Tr}\left[ \gamma _{5}S_{b}^{bb^{\prime }}(x)\gamma
_{5}S_{c}^{b^{\prime }b}(-x)\right] ,  \label{eq:QCD1}
\end{eqnarray}%
where propagators of the $b$ and $c$-quarks are denoted by $S_{b(c)}(x)$
\cite{Agaev:2020zad}, respectively.

In the case under consideration $\Pi ^{\mathrm{OPE}}(p)$ is equal to the
invariant amplitude $\Pi ^{\mathrm{OPE}}(p^{2})$. To find SRs for the $m$
and $\Lambda $ one equates $\Pi ^{\mathrm{OPE}}(p^{2})$ and $\Pi ^{\mathrm{%
Phys}}(p^{2})$, uses the Borel transformation and, benefiting from the
quark-hadron duality assumption, carries out continuum subtraction \cite%
{Shifman:1978bx,Shifman:1978by}. As a result, one obtains
\begin{equation}
m^{2}=\frac{\Pi ^{\prime }(M^{2},s_{0})}{\Pi (M^{2},s_{0})},  \label{eq:Mass}
\end{equation}%
with $\Pi ^{\prime }(M^{2},s_{0})=\frac{d\Pi
(M^{2},s_{0})}{d(-1/M^{2})}$ and
\begin{equation}
\Lambda ^{2}=e^{m^{2}/M^{2}}\Pi (M^{2},s_{0}),  \label{eq:Coupl}
\end{equation}%
which are the SRs for the parameters of $\mathcal{M}$. Above $\Pi
(M^{2},s_{0})$ is the amplitude $\Pi ^{\mathrm{OPE}}(p^{2})$ after the Borel
transformation and continuum subtraction, and it depends on Borel and continuum
subtraction parameters $M^{2}$ and $s_{0}$. 

The amplitude $\Pi (M^{2},s_{0})$ is a sum of the integral of the spectral
density $\rho (s)$ and the term $\Pi (M^{2})$
\begin{eqnarray}
\Pi (M^{2},s_{0}) &=&\int_{4\mathcal{(}m_{b}+m_{c})^{2}}^{s_{0}}ds\rho
(s)e^{-s/M^{2}}+\Pi (M^{2}).  \notag \\
&&  \label{eq:InvAmp}
\end{eqnarray}%
The function $\rho (s)$ is equal to the imaginary part of $\Pi ^{\mathrm{OPE}%
}(p^{2})$, whereas $\Pi (M^{2})$ is evaluated directly from $\Pi ^{\mathrm{%
OPE}}(p^{2})$ and does not contain terms included into $\rho (s)$. We do not
write down explicitly $\rho (s)$ and $\Pi (M^{2})$ because they are given by
lengthy formulas.

To perform numerical computations, one needs to specify parameters which
enter to Eqs.\ (\ref{eq:Mass}) and (\ref{eq:Coupl}). The masses of the
quarks $m_{b}$ and\ $m_{c}$, and gluon condensate $\langle \alpha
_{s}G^{2}/\pi \rangle $ are universal quantities which do not depend on a
considering problem:%
\begin{eqnarray}
&&m_{b}=(4.183\pm 0.007)\ \mathrm{GeV}\text{,\ }m_{c}=(1.2730\pm 0.0046)~%
\mathrm{GeV}\text{,}  \notag \\
&&\langle \alpha _{s}G^{2}/\pi \rangle =(0.012\pm 0.004)~\mathrm{GeV}^{4}.
\label{eq:GluonCond}
\end{eqnarray}%
In Eq.\ (\ref{eq:GluonCond})\ $m_{b}$ and $m_{c}$ are the running
(scale-dependent) $b$ and $c$ quark masses $\overline{m}_{b}$ and $\overline{%
m}_{c}$ in the $\overline{\mathrm{MS}}$ scheme at the scales $\mu =\overline{%
m}_{b}$ and $\mu =\overline{m}_{c}$ \cite{PDG:2024}, respectively. Note that
there are various mass definitions, i.e., schemes for the heavy quarks (see,
the review "Quark Masses" in Ref. \cite{PDG:2024}). The masses of quarks
expressed in some of such schemes can be converted to another one by means
of known formulas. The $\overline{\mathrm{MS}}$ scheme is appropriate for
our purposes, because it effectively suppresses nonperturbative effects
ensuring convergence of $\mathrm{OPE}$ and improves precision of
calculations. The gluon condensate $\langle \alpha _{s}G^{2}/\pi \rangle $
is $\mu $-scale independent parameter: Its numerical value was extracted
from analysis of hadronic processes \cite{Shifman:1978bx,Shifman:1978by}.

In contrast, parameters $M^{2}$ and $s_{0}$ depend on a task under analysis and
their choice should meet requirements of the SR calculations. Namely, they
have to be fixed in such a way that to ensure $\mathrm{PC}\geq 0.5$, where $%
\mathrm{PC}$ is the pole contribution defined as
\begin{equation}
\mathrm{PC}=\frac{\Pi (M^{2},s_{0})}{\Pi (M^{2},\infty )}.  \label{eq:PC}
\end{equation}%
The convergence of the operator product expansion and stability of extracted
quantities $m$ and $\Lambda $ on these parameters are also among necessary
conditions. In present computations we take into account the dimension-$4$
nonperturbative term. Therefore to guarantee convergence of $\mathrm{OPE}$
it is enough to impose a constraint on this term. We restrict this term by
the requirement
\begin{equation}
|\Pi ^{\mathrm{Dim4}}(M^{2},s_{0})|\leq 0.05|\Pi (M^{2},s_{0})|.
\label{eq:OPE}
\end{equation}%
Note that the upper border of $M^{2}$ is fixed using Eq.\ (\ref{eq:PC}). The
constraint Eq.\ (\ref{eq:OPE}) allows one to fix low limit of the Borel
parameter.

\begin{figure}[h]
\includegraphics[width=8.5cm]{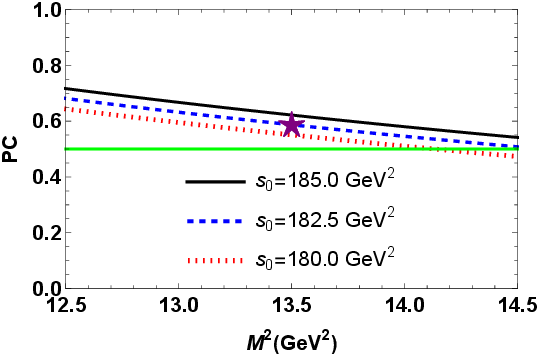}
\caption{The pole contribution $\mathrm{PC}$ as a function of $M^{2}$ at
fixed $s_{0}$. The red star marks the point $M^{2}=13.5~\mathrm{GeV}^{2}$
and $s_{0}=182.5~\mathrm{GeV}^{2}$. }
\label{fig:PC}
\end{figure}

Our calculations prove that the working windows
\begin{eqnarray}
M^{2} &\in &[12.5,14.5]~\mathrm{GeV}^{2},\ s_{0}\in \lbrack 180,185]~\mathrm{%
GeV}^{2}.  \notag \\
&&  \label{eq:Wind1}
\end{eqnarray}%
satisfy these constraints. Indeed, at $M^{2}=14.5~\mathrm{GeV}^{2}$ one gets
$\mathrm{PC}\approx 0.51$ (on the average in $s_{0})$, whereas at $%
M^{2}=12.5~\mathrm{GeV}^{2}$ the pole contribution is equal to $\mathrm{PC}%
\approx 0.68$. The term $|\Pi ^{\mathrm{Dim4}}(M^{2},s_{0})|$ at $M^{2}=12.5~%
\mathrm{GeV}^{2}$ does not exceed $1\%$ of the whole result. In Fig.\ \ref%
{fig:PC}, we plot dependence of $\mathrm{PC}$ on the parameter $M^{2}$ at
some values of $s_{0}$. It is clear, that\ $\mathrm{PC}$ overshoots $0.5$
for almost all $M^{2}$ and $s_{0}$ from the working windows.

\begin{widetext}

\begin{figure}[h!]
\begin{center}
\includegraphics[totalheight=6cm,width=8cm]{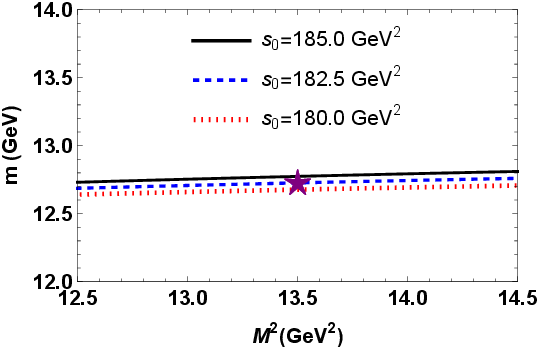}
\includegraphics[totalheight=6cm,width=8cm]{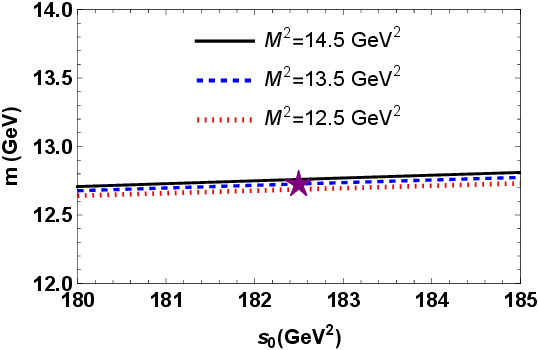}
\end{center}
\caption{Mass $m$ of the hadronic molecule $\mathcal{M}$ vs the parameters $M^{2}$ (left panel), and $s_0$ (right panel).}
\label{fig:Mass}
\end{figure}

\end{widetext}

The spectral parameters of the molecule $\mathcal{M}$ are computed as their
average values in the domains Eq.\ (\ref{eq:Wind1}), and amount to
\begin{eqnarray}
m &=&(12725\pm 85)~\mathrm{MeV},  \notag \\
\Lambda &=&(1.40\pm 0.16)~\mathrm{GeV}^{5}.  \label{eq:Result1}
\end{eqnarray}%
These results coincide with the sum rule predictions at the point $%
M^{2}=13.5~\mathrm{GeV}^{2}$ and $s_{0}=182.5~\mathrm{GeV}^{2}$, where the
pole term forms $\mathrm{PC}\approx 59\%$ of the whole contribution. This
ensures the prevalence of the pole contribution in the obtained results, and
shows ground-state character of the molecule $\mathcal{M}$ in its class. As the function of the parameters $M^{2}$ and $s_{0}$ the mass of the molecule $\mathcal{M}$ is depicted in Fig.\ \ref{fig:Mass}.

\section{Dissociations of $\mathcal{M}$ to conventional mesons}

\label{sec:Widths1}


In this section, we consider dissociations of the scalar molecule $\mathcal{M%
}$ into pairs of ordinary mesons and calculate partial widths of these
channels. In these processes constituent quarks of $\mathcal{M}$ form
directly the final-state particles. The mass $m$ and spin-parities $J^{%
\mathrm{PC}}=0^{++}$ of $\mathcal{M}$ are key parameters that limit its
possible decay channels. In fact, it is evident that the mass of $\mathcal{M}
$ should exceed the corresponding two-meson thresholds. It is easy to see
that decays to mesons $\eta _{b}\eta _{c}$, $J/\psi \Upsilon $, and $%
B_{c}^{+}B_{c}^{-}$ meet this requirement. The quantum numbers of the
final-state mesons are also compatible with ones of the hadronic molecule $%
\mathcal{M}$.


\subsection{$\mathcal{M}\rightarrow \protect\eta _{b}\protect\eta _{c}$}


We first explore the process $\mathcal{M}\rightarrow \eta _{b}\eta _{c}$.
The width of this decay, besides the masses and decay constants (current
couplings) of the initial and final-state particles, contains the coupling $%
g_{1}$ which characterizes their strong interaction at the vertex $\mathcal{M%
}\eta _{b}\eta _{c}$. To evaluate $g_{1}$, we consider the QCD correlator
\begin{eqnarray}
\Pi _{1}(p,p^{\prime }) &=&i^{2}\int d^{4}xd^{4}ye^{ip^{\prime
}y}e^{-ipx}\langle 0|\mathcal{T}\{J^{\eta _{b}}(y)  \notag \\
&&\times J^{\eta _{c}}(0)J^{\dagger }(x)\}|0\rangle ,  \label{eq:CF3}
\end{eqnarray}%
with
\begin{equation}
J^{\eta _{b}}(x)=\overline{b}_{i}(x)i\gamma _{5}b_{i}(x),\ J^{\eta _{c}}(x)=%
\overline{c}_{j}(x)i\gamma _{5}c_{j}(x),  \label{eq:CR3}
\end{equation}%
being the pseudoscalar $\eta _{b}$ and $\eta _{c}$ mesons' currents,
respectively.

Analysis of this correlation function enables us to obtain the SR for the
form factor $g_{1}(q^{2})$, which at the mass shell $q^{2}=m_{\eta _{c}}^{2}$
gives $g_{1}$. To get the sum rule for $g_{1}(q^{2})$, we recast $\Pi
_{1}(p,p^{\prime })$ into the form
\begin{eqnarray}
&&\Pi _{1}^{\mathrm{Phys}}(p,p^{\prime })=\frac{\langle 0|J^{\eta _{b}}|\eta
_{b}(p^{\prime })\rangle }{p^{\prime 2}-m_{\eta _{b}}^{2}}\frac{\langle
0|J^{\eta _{c}}|\eta _{c}(q)\rangle }{q^{2}-m_{\eta _{c}}^{2}}  \notag \\
&&\times \langle \eta _{b}(p^{\prime })\eta _{c}(q)|\mathcal{M}(p)\rangle
\frac{\langle \mathcal{M}(p)|J^{\dagger }|0\rangle }{p^{2}-m^{2}}+\cdots ,
\notag \\
&&  \label{eq:CF5}
\end{eqnarray}%
where $m_{\eta _{b}}=(9398.7\pm 2.0)~\mathrm{MeV}$ and $m_{\eta
_{c}}=(2983.9\pm 0.4)~\mathrm{MeV}$ are the masses of the quarkonia $\eta
_{b}$ and $\eta _{c}$ \cite{PDG:2024}, respectively. The term in the
correlation function $\Pi _{1}^{\mathrm{Phys}}(p,p^{\prime })$ is a
contribution of the ground-state particles, whereas the dots denote effects
of higher resonances and continuum states. Next, we introduce the matrix
elements
\begin{equation}
\langle 0|J^{\eta _{b}}|\eta _{b}(p^{\prime })\rangle =\frac{f_{\eta
_{b}}m_{\eta _{b}}^{2}}{2m_{b}},\ \langle 0|J^{\eta _{c}}|\eta
_{c}(q)\rangle =\frac{f_{\eta _{c}}m_{\eta _{c}}^{2}}{2m_{c}},
\label{eq:ME2}
\end{equation}%
where $f_{\eta _{b}}=724~\mathrm{MeV}$ and $f_{\eta _{c}}=(421\pm 35)~%
\mathrm{MeV}$ are decay constants of the mesons $\eta _{b}$ and $\eta _{c}$.

The vertex $\mathcal{M}\eta _{b}\eta _{c}$ is given by the formula%
\begin{equation}
\langle \eta _{b}(p^{\prime })\eta _{c}(q)|\mathcal{M}(p)\rangle
=g_{1}(q^{2})p\cdot p^{\prime }.  \label{eq:ME3}
\end{equation}%
Having performed all calculations, for $\Pi _{1}^{\mathrm{Phys}}(p,p^{\prime
})$ we get
\begin{eqnarray}
&&\Pi _{1}^{\mathrm{Phys}}(p,p^{\prime })=g_{1}(q^{2})\frac{\Lambda f_{\eta
_{b}}m_{\eta _{b}}^{2}f_{\eta _{c}}m_{\eta _{c}}^{2}}{8m_{b}m_{c}\left(
p^{2}-m^{2}\right) }  \notag \\
&&\times \frac{m^{2}+m_{\eta _{b}}^{2}-q^{2}}{\left( p^{\prime 2}-m_{\eta
_{b}}^{2}\right) (q^{2}-m_{\eta _{c}}^{2})}+\cdots .  \label{eq:CF6}
\end{eqnarray}%
The correlator $\Pi _{1}^{\mathrm{Phys}}(p,p^{\prime })$ has a simple
Lorentz structure, therefore it is equal to the invariant amplitude $\Pi
_{1}^{\mathrm{Phys}}(p^{2},p^{\prime 2},q^{2})$.

The same correlation function $\Pi _{1}(p,p^{\prime })$ computed in terms of
the heavy quark propagators reads
\begin{eqnarray}
&&\Pi _{1}^{\mathrm{OPE}}(p,p^{\prime })=i^{4}\int d^{4}xd^{4}ye^{ip^{\prime
}y}e^{-ipx}\mathrm{Tr}\left[ \gamma _{5}S_{b}^{ib}(y-x)\right.  \notag \\
&&\left. \times \gamma _{5}S_{c}^{bj}(x)\gamma _{5}S_{c}^{ja}(-x)\gamma
_{5}S_{b}^{ai}(x-y)\right] .  \label{eq:QCDside2}
\end{eqnarray}%
Due to a triviality of its structure $\Pi _{1}^{\mathrm{OPE}}(p,p^{\prime })$
is equal to the amplitude $\Pi _{1}^{\mathrm{OPE}}(p^{2},p^{\prime 2},q^{2})$%
. Having equated the amplitudes $\Pi _{1}^{\mathrm{Phys}}(p^{2},p^{\prime
2},q^{2})$ and $\Pi _{1}^{\mathrm{OPE}}(p^{2},p^{\prime 2},q^{2})$,
performed the double Borel transformations over $-p^{2}$, $-p^{\prime 2}$
and continuum subtractions, one gets the sum rule for $g_{1}(q^{2})$%
\begin{eqnarray}
&&g_{1}(q^{2})=\frac{8m_{b}m_{c}}{\Lambda f_{\eta _{b}}m_{\eta
_{b}}^{2}f_{\eta _{c}}m_{\eta _{c}}^{2}}\frac{q^{2}-m_{\eta _{c}}^{2}}{%
m^{2}+m_{\eta _{b}}^{2}-q^{2}}  \notag \\
&&\times e^{m^{2}/M_{1}^{2}}e^{m_{\eta _{b}}^{2}/M_{2}^{2}}\Pi _{1}(\mathbf{M%
}^{2},\mathbf{s}_{0},q^{2}).  \label{eq:SRCoup2}
\end{eqnarray}%
In Eq.\ (\ref{eq:SRCoup2}) the correlator $\Pi _{1}(\mathbf{M}^{2},\mathbf{s}%
_{0},q^{2})$ has the following form
\begin{eqnarray}
&&\Pi _{1}(\mathbf{M}^{2},\mathbf{s}_{0},q^{2})=%
\int_{4(m_{b}+m_{c})^{2}}^{s_{0}}ds\int_{4m_{b}^{2}}^{s_{0}^{\prime
}}ds^{\prime }\rho _{1}(s,s^{\prime },q^{2})  \notag \\
&&\times e^{-s/M_{1}^{2}}e^{-s^{\prime }/M_{2}^{2}}.
\end{eqnarray}%
It is written using the spectral density $\rho _{1}(s,s^{\prime },q^{2})$
which contains the perturbative and dimension-$4$ components. The function $%
\Pi _{1}(\mathbf{M}^{2},\mathbf{s}_{0},q^{2})$ depends on the Borel $\mathbf{%
M}^{2}=(M_{1}^{2},M_{2}^{2})$ and continuum threshold parameters $\mathbf{s}%
_{0}=(s_{0},s_{0}^{\prime })$. The pair of parameters $(M_{1}^{2},s_{0})$
corresponds to the hadronic $\mathcal{M}$ molecule channel, whereas $%
(M_{2}^{2},s_{0}^{\prime })$ describes the channel of the $\eta _{b}$ meson.

To perform numerical analysis, in the $\mathcal{M}$ channel we employ $%
M_{1}^{2}$ and $s_{0}$ from Eq.\ (\ref{eq:Wind1}). In the $\eta _{b}$
channel as the parameters $(M_{2}^{2},\ s_{0}^{\prime })$ we use
\begin{equation}
M_{2}^{2}\in \lbrack 9,11]~\mathrm{GeV}^{2},\ s_{0}^{\prime }\in \lbrack
95,99]~\mathrm{GeV}^{2}.  \label{eq:Wind3}
\end{equation}%
Note that$\ \sqrt{s_{0}^{\prime }}$ is restricted by the mass $9.999\
\mathrm{GeV}$ of the first radially excited $\eta _{b}(2S)$ meson.

The credible results for the form factor $g_{1}(q^{2})$ in the sum rule
context is obtained in the Euclidean region $q^{2}<0$. But the coupling $%
g_{1}$ has to be determined at $q^{2}=m_{\eta _{c}}^{2}$. This problem can
be avoided by introducing a variable $Q^{2}=-q^{2}$ and employing the
notation $g_{1}(Q^{2})$ for the new function. Afterwards, we employ an
extrapolating function $\mathcal{G}_{1}(Q^{2})$ which for $Q^{2}>0$
coincides with the SR data, but can also be computed at $Q^{2}<0$. For these
purposes, we use the function $\mathcal{G}_{i}(Q^{2})$
\begin{equation}
\mathcal{G}_{i}(Q^{2})=\mathcal{G}_{i}^{0}\mathrm{\exp }\left[ a_{i}^{1}%
\frac{Q^{2}}{m^{2}}+a_{i}^{2}\left( \frac{Q^{2}}{m^{2}}\right) ^{2}\right] ,
\end{equation}%
where $\mathcal{G}_{i}^{0}$, $a_{i}^{1}$, and $a_{i}^{2}$ are parameters
extracted from comparison with the SR data.

\begin{figure}[h]
\includegraphics[width=8.5cm]{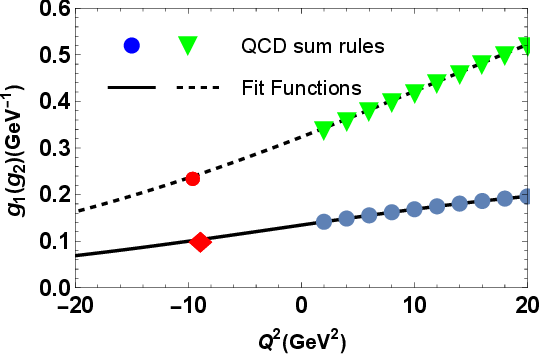}
\caption{QCD data and fit functions for the form factor $g_{1}(Q^{2})$
(solid line) and $g_{2}(Q^{2})$ (dashed line). The diamond and circle fix
the points $Q^{2}=-m_{\protect\eta _{c}}^{2}$ and $Q^{2}=-m_{J/\protect\psi%
}^{2}$, respectively. }
\label{fig:Fit}
\end{figure}

In this article, the SR computations are carried out for $Q^{2}=2-20~\mathrm{%
GeV}^{2}$. Relevant numerical predictions for $g_{1}(Q^{2})$ are depicted in
Fig.\ \ref{fig:Fit}. Then, one easily determines the parameters $\mathcal{G}%
_{1}^{0}=0.13~\mathrm{GeV}^{-1}$, $a_{1}^{1}=4.24$, and $a_{1}^{2}=-9.62$ of
the function $\mathcal{G}_{1}(Q^{2})$: This function is shown in Fig.\ \ref%
{fig:Fit}, as well.

Finally, we find $g_{1}$
\begin{equation}
g_{1}\equiv \mathcal{G}_{1}(-m_{\eta _{c}}^{2})=(1.0\pm 0.2)\times 10^{-1}\
\mathrm{GeV}^{-1}.
\end{equation}%
We calculate the width of the decay $\mathcal{M}\rightarrow \eta _{b}\eta
_{c}$ by means of the formula
\begin{equation}
\Gamma \left[ \mathcal{M}\rightarrow \eta _{b}\eta _{c}\right] =g_{1}^{2}%
\frac{m_{\eta _{b}}^{2}\lambda _{1}}{8\pi }\left( 1+\frac{\lambda _{1}^{2}}{%
m_{\eta _{b}}^{2}}\right) ,  \label{eq:PDw2}
\end{equation}%
where $\lambda _{1}=\lambda (m,m_{\eta _{b}},m_{\eta _{c}})$, and
\begin{equation}
\lambda (x,y,z)=\frac{\sqrt{%
x^{4}+y^{4}+z^{4}-2(x^{2}y^{2}+x^{2}z^{2}+y^{2}z^{2})}}{2x}.
\end{equation}%
As a result, we obtain
\begin{equation}
\Gamma \left[ \mathcal{M}\rightarrow \eta _{b}\eta _{c}\right] =(48.4\pm
15.0)~\mathrm{MeV}.  \label{eq:DW1}
\end{equation}

A part $\pm 13.7$ of the total errors in Eq.\ (\ref{eq:DW1}) is connected
with the uncertainties of the coupling $g_{1}$, whereas $\pm 6.1$ generated
by the ambiguities of the particles' masses in Eq.\ (\ref{eq:PDw2}).


\subsection{$\mathcal{M}\rightarrow J/\protect\psi \Upsilon $}


Here, we consider the decay $\mathcal{M}\rightarrow J/\psi \Upsilon $ and
extract the strong coupling $g_{2}$ at the vertex $\mathcal{M}J/\psi
\Upsilon $ which is equal to the form factor $g_{2}(q^{2})$ at the mass
shell $q^{2}=m_{J/\psi }^{2}$.

The SR for the form factor $g_{2}(q^{2})$ is derived by means of the
correlation function
\begin{eqnarray}
\Pi _{2\mu \nu }(p,p^{\prime }) &=&i^{2}\int d^{4}xd^{4}ye^{ip^{\prime
}y}e^{-ipx}\langle 0|\mathcal{T}\{J_{\mu }^{\Upsilon }(y)  \notag \\
&&\times J_{\nu }^{J/\psi }(0)J^{\dagger }(x)\}|0\rangle ,  \label{eq:CF1a}
\end{eqnarray}%
where $J_{\mu }^{\Upsilon }(x)$ and $J_{\nu }^{J/\psi }(x)$ are
interpolating currents of the vector quarkonia $\Upsilon $ and $J/\psi $,
respectively. They are defined as
\begin{equation}
J_{\mu }^{\Upsilon }(x)=\overline{b}_{i}(x)\gamma _{\mu }b_{i}(x),\ J_{\nu
}^{J/\psi }(x)=\overline{c}_{j}(x)\gamma _{\nu }c_{j}(x),
\end{equation}%
with $i$ and $j$ being the color indices.

To find the phenomenological component $\Pi _{2\mu \nu }^{\mathrm{Phys}%
}(p,p^{\prime })$ of the sum rule, we need to rewrite Eq.\ (\ref{eq:CF1a})
using the involved particles' physical parameters. By taking into account
only contributions of the ground-level states, we recast the correlator $\Pi
_{2\mu \nu }(p,p^{\prime })$ into the form
\begin{eqnarray}
&&\Pi _{2\mu \nu }^{\mathrm{Phys}}(p,p^{\prime })=\frac{\langle 0|J_{\mu
}^{\Upsilon }|\Upsilon (p^{\prime },\varepsilon _{1})\rangle }{p^{\prime
2}-m_{\Upsilon }^{2}}\frac{\langle 0|J_{\nu }^{J/\psi }|J/\psi
(q,\varepsilon _{2})\rangle }{q^{2}-m_{J/\psi }^{2}}  \notag \\
&&\times \langle \Upsilon (p^{\prime },\varepsilon _{1})J/\psi
(q,\varepsilon _{2})|\mathcal{M}(p\rangle \frac{\langle \mathcal{M}%
(p)|J^{\dagger }|0\rangle }{p^{2}-m^{2}}+\cdots ,  \notag \\
&&  \label{eq:TP1}
\end{eqnarray}%
where $m_{\Upsilon }=(9460.40\pm 0.09\pm 0.04)~\mathrm{MeV}$ and $m_{J/\psi
}=(3096.900\pm 0.006)~\mathrm{MeV}$ are masses of the $\Upsilon $ and $%
J/\psi $ mesons \cite{PDG:2024}. In the expression above, we denote by $%
\varepsilon _{1}$ and $\varepsilon _{2}$ the polarization vectors of these
quarkonia, respectively.

To further simplify Eq.\ (\ref{eq:TP1}), it is convenient to employ the
matrix elements of the mesons $\Upsilon $ and $J/\psi $%
\begin{eqnarray}
\langle 0|J_{\mu }^{\Upsilon }|\Upsilon (p^{\prime },\varepsilon
_{1})\rangle &=&f_{\Upsilon }m_{\Upsilon }\varepsilon _{1\mu },  \notag \\
\langle 0|J_{\nu }^{J/\psi }|J/\psi (q,\varepsilon _{2})\rangle &=&f_{J/\psi
}m_{J/\psi }\varepsilon _{2\nu }.  \label{eq:C2}
\end{eqnarray}%
Here, $f_{\Upsilon }=(708\pm 8)~\mathrm{MeV}$ and $f_{J/\psi }=(411\pm 7)~%
\mathrm{MeV}$ are decay constants of the mesons: Their experimental values
are borrowed from Ref.\ \cite{Lakhina:2006vg}.

Besides, one should specify the matrix element $\langle \Upsilon (p^{\prime
},\varepsilon _{1})J/\psi (q,\varepsilon _{2})|\mathcal{M}(p)\rangle $ which
depends on the momenta and polarization vectors of the particles $\mathcal{M}
$, $\Upsilon $ and $J/\psi $
\begin{eqnarray}
&&\langle \Upsilon (p^{\prime },\varepsilon _{1})J/\psi (q,\varepsilon _{2})|%
\mathcal{M}(p)\rangle =g_{2}(q^{2})\left[ q\cdot p^{\prime }\right.  \notag
\\
&&\left. \times \varepsilon _{1}^{\ast }\cdot \varepsilon _{2}^{\ast
}-q\cdot \varepsilon _{1}^{\ast }p^{\prime }\cdot \varepsilon _{2}^{\ast }
\right] .
\end{eqnarray}%
As a result, for $\Pi _{2\mu \nu }^{\mathrm{Phys}}(p,p^{\prime })$ we get
the expression
\begin{eqnarray}
&&\Pi _{2\mu \nu }^{\mathrm{Phys}}(p,p^{\prime })=g_{2}(q^{2})\frac{\Lambda
f_{\Upsilon }m_{\Upsilon }f_{J/\psi }m_{J/\psi }}{\left( p^{2}-m^{2}\right)
(p^{\prime 2}-m_{\Upsilon }^{2})(q^{2}-m_{J/\psi }^{2})}  \notag \\
&&\times \left[ \frac{(m^{2}-m_{\Upsilon }^{2}-q^{2})}{2}g_{\mu \nu }-p_{\nu
}^{\prime }q_{\mu }\right] +\cdots .
\end{eqnarray}

The function $\Pi _{2\mu \nu }^{\mathrm{OPE}}(p,p^{\prime })$, i.e., the QCD
component of the sum rule is given by the expression
\begin{eqnarray}
&&\Pi _{2\mu \nu }^{\mathrm{OPE}}(p,p^{\prime })=\int
d^{4}xd^{4}ye^{ip^{\prime }y}e^{-ipx}\mathrm{Tr}\left[ \gamma _{\mu
}S_{b}^{ib}(y-x)\right.  \notag \\
&&\left. \times \gamma _{5}S_{c}^{bj}(x)\gamma _{\nu }S_{c}^{ja}(-x)\gamma
_{5}S_{b}^{ai}(x-y)\right] .
\end{eqnarray}%
We utilize the structures proportional to $g_{\mu \nu }$ in the correlators
and use corresponding amplitudes $\Pi _{2}^{\mathrm{Phys}}(p^{2},p^{\prime
2},q^{2})$ and $\Pi _{2}^{\mathrm{OPE}}(p^{2},p^{\prime 2},q^{2})$ to find
SR for the form factor $g_{2}(q^{2})$. After standard operations the sum
rule for $g_{2}(q^{2})$ reads
\begin{eqnarray}
g_{2}(q^{2}) &=&\frac{2(q^{2}-m_{J/\psi }^{2})}{\Lambda f_{\Upsilon
}m_{\Upsilon }f_{J/\psi }m_{J/\psi }(m^{2}-m_{\Upsilon }^{2}-q^{2})}  \notag
\\
&&\times e^{m^{2}/M_{1}^{2}}e^{m_{\Upsilon }^{2}/M_{2}^{2}}\Pi _{2}(\mathbf{M%
}^{2},\mathbf{s}_{0},q^{2}).  \label{eq:SRG}
\end{eqnarray}

Requirements which should be satisfied by the auxiliary parameters $\mathbf{M%
}^{2}$ and $\mathbf{s}_{0}$ are universal for all SR computations and have
been explained in the previous section. Numerical analysis shows that the
regions in Eq.\ (\ref{eq:Wind1}) for the parameters $(M_{1}^{2},s_{0})$ and
\begin{equation}
M_{2}^{2}\in \lbrack 10,12]~\mathrm{GeV}^{2},\ s_{0}^{\prime }\in \lbrack
98,100]~\mathrm{GeV}^{2}.  \label{eq:Wind3A}
\end{equation}%
for $(M_{2}^{2},s_{0}^{\prime })$ satisfy all these requirements. Note that $%
s_{0}^{\prime }$ is limited by the mass $m_{\Upsilon (2S)}=(10023.4\pm 0.5)~%
\mathrm{MeV}\ $of the radially excited state $\Upsilon (2S)$, i.e., $%
s_{0}^{\prime }<m_{\Upsilon (2S)}^{2}$.

We calculate $g_{2}(Q^{2})$ in the domain $Q^{2}=2-20~\mathrm{GeV}^{2}$ and
fit these data by the function $\mathcal{G}_{2}(Q^{2})$ with parameters $%
\mathcal{G}_{1}^{0}=0.32~\mathrm{GeV}^{-1}$, $a_{1}^{1}=4.71$, and $%
a_{1}^{2}=-6.81.$ The coupling $g_{2}$ is extracted at $q^{2}=m_{J/\psi
}^{2} $, i.e., at $Q^{2}=-m_{J/\psi }^{2}$
\begin{equation}
g_{2}\equiv \mathcal{G}_{2}(-m_{J/\psi }^{2})=(2.4\pm 0.5)\times 10^{-1}\
\mathrm{GeV}^{-1}.  \label{eq:g1}
\end{equation}

The partial width of the decay $\mathcal{M}\rightarrow J/\psi \Upsilon $ is
determined by the expression%
\begin{eqnarray}
&&\Gamma \left[ \mathcal{M}\rightarrow J/\psi \Upsilon \right] =g_{2}^{2}%
\frac{\lambda _{2}}{16\pi m^{2}}\left[ m^{4}+m_{\Upsilon }^{4}+m_{J/\psi
}^{4}\right. ,  \notag \\
&&\left. +4m_{\Upsilon }^{2}m_{J/\psi }^{2}-2m^{2}(m_{\Upsilon
}^{2}+m_{J/\psi }^{2})\right] .  \label{eq:PDw2A}
\end{eqnarray}%
where $\lambda _{2}=\lambda (m,m_{\Upsilon },m_{J/\psi })$. Our computations
yield

\begin{equation}
\Gamma \left[ \mathcal{M}\rightarrow J/\psi \Upsilon \right] =(35.3\pm 14.4)~%
\mathrm{MeV}.
\end{equation}


\subsection{$\mathcal{M}\rightarrow B_{c}^{+}B_{c}^{-}$}


The partial width of the decay $\mathcal{M}\rightarrow B_{c}^{+}B_{c}^{-}$
is fixed by the coupling $g_{3}$ at the vertex $\mathcal{M}%
B_{c}^{+}B_{c}^{-} $. In the framework of the QCD SR method the form factor $%
g_{3}(q^{2})$ is calculated using the three-point correlation function%
\begin{eqnarray}
&&\Pi _{3}(p,p^{\prime })=i^{2}\int d^{4}xd^{4}ye^{ip^{\prime
}y}e^{-ipx}\langle 0|\mathcal{T}\{J^{B_{c}^{+}}(y)  \notag \\
&&\times J^{B_{c}^{-}}(0)J^{\dagger }(x)\}|0\rangle ,  \label{eq:CF7}
\end{eqnarray}%
where $J^{B_{c}^{+}}(x)$ and $J^{B_{c}^{-}}(x)$ are the interpolating
currents for the mesons $B_{c}^{+}$ and $B_{c}^{-}$
\begin{equation}
J^{B_{c}^{+}}(x)=\overline{b}_{i}(x)i\gamma _{5}c_{i}(x),\ J^{B_{c}^{-}}(x)=%
\overline{c}_{j}(x)i\gamma _{5}b_{j}(x).
\end{equation}

To compute the physical part of the SR, we make use of the expressions
\begin{equation}
\langle 0|J^{B_{c}^{\pm }}|B_{c}^{\pm }\rangle =\frac{f_{B_{c}}m_{B_{c}}^{2}%
}{m_{b}+m_{c}},  \label{eq:ME2A}
\end{equation}%
and
\begin{equation}
\langle B_{c}^{+}(p^{\prime })B_{c}^{-}(q)|\mathcal{M}\rangle
=g_{3}(q^{2})p\cdot p^{\prime },  \label{eq:ME3A}
\end{equation}%
where $m_{B_{c}}=(6274.47\pm 0.27\pm 0.17)~\mathrm{MeV}$ and $%
f_{B_{c}}=(371\pm 37)~\mathrm{MeV}$ are the parameters of the mesons $%
B_{c}^{\pm }$ \cite{PDG:2024,Wang:2024fwc}. In terms of $m_{B_{c}}$ and $%
f_{B_{c}}$ the correlation function $\Pi _{3}(p,p^{\prime })$ acquires the
form%
\begin{eqnarray}
&&\Pi _{3}^{\mathrm{Phys}}(p,p^{\prime })=\frac{g_{3}(q^{2})\Lambda
f_{B_{c}}^{2}m_{B_{c}}^{4}}{(m_{b}+m_{c})^{2}\left( p^{2}-m^{2}\right)
\left( p^{\prime 2}-m_{B_{c}}^{2}\right) }  \notag \\
&&\times \frac{m^{2}+m_{B_{c}}^{2}-q^{2}}{2(q^{2}-m_{B_{c}}^{2})}+\cdots .
\label{eq:CR2A}
\end{eqnarray}

The QCD component of SR for the form factor $g_{3}(q^{2})$ is given by the
expression%
\begin{eqnarray}
&&\Pi _{3}^{\mathrm{OPE}}(p,p^{\prime })=-i^{4}\int
d^{4}xd^{4}ye^{ip^{\prime }y}e^{-ipx}\mathrm{Tr}\left[ \gamma
_{5}S_{c}^{ia}(y-x)\right.  \notag \\
&&\left. \times \gamma _{5}S_{b}^{ai}(x-y)\right] \mathrm{Tr}\left[ \gamma
_{5}S_{c}^{bj}(x)\gamma _{5}S_{b}^{jb}(-x)\right] .  \label{eq:QCDside}
\end{eqnarray}%
In the process under consideration, the correlators $\Pi _{3}^{\mathrm{Phys}%
}(p,p^{\prime })$ and $\Pi _{3}^{\mathrm{OPE}}(p,p^{\prime })$ have simple
Lorentz organizations. Having denoted by $\Pi _{3}^{\mathrm{Phys}%
}(p^{2},p^{\prime 2},q^{2})$ and $\Pi _{3}^{\mathrm{OPE}}(p^{2},p^{\prime
2},q^{2})$ corresponding amplitudes, we find the sum rule

\begin{eqnarray}
&&g_{3}(q^{2})=\frac{2(m_{b}+m_{c})^{2}}{\Lambda f_{B_{c}}^{2}m_{B_{c}}^{4}}%
\frac{q^{2}-m_{B_{c}}^{2}}{m^{2}+m_{B_{c}}^{2}-q^{2}}  \notag \\
&&\times e^{m^{2}/M_{1}^{2}}e^{m_{B_{c}}^{2}/M_{2}^{2}}\Pi _{3}(\mathbf{M}%
^{2},\mathbf{s}_{0},q^{2}).  \label{eq:SRCoup}
\end{eqnarray}%
Here, $\Pi _{3}(\mathbf{M}^{2},\mathbf{s}_{0},q^{2})$ is the Borel
transformed and subtracted amplitude $\Pi _{3}^{\mathrm{OPE}%
}(p^{2},p^{\prime 2},q^{2})$.

The remaining operations are similar to ones explained above. Therefore, we
present only the windows for $M_{2}^{2}$, and $s_{0}^{\prime }$ in the $%
B_{c}^{+}$ meson channel
\begin{equation}
M_{2}^{2}\in \lbrack 6.5,7.5]~\mathrm{GeV}^{2},\ s_{0}^{\prime }\in \lbrack
45,47]~\mathrm{GeV}^{2}.
\end{equation}%
The function $\mathcal{G}_{3}(Q^{2})$ has the parameters: $\mathcal{G}%
_{3}^{0}=0.21~\mathrm{GeV}^{-1}$, $a_{3}^{1}=1.79$, and $a_{3}^{2}=-4.59$.
Then, the strong coupling $g_{3}$ is equal to
\begin{equation}
g_{3}\equiv \mathcal{G}_{3}(-m_{B_{c}}^{2})=(1.0\pm 0.2)\times 10^{-1}\
\mathrm{GeV}^{-1}.
\end{equation}%
Having employed $g_{3}$, and Eq.\ (\ref{eq:PDw2}) with evident replacements $%
m_{\eta _{b}}\rightarrow m_{B_{c}}$ and $\lambda _{3}\rightarrow \lambda
_{3}=\lambda (m,m_{B_{c}},m_{B_{c}})$, we find
\begin{equation}
\Gamma \left[ \mathcal{M}\rightarrow B_{c}^{+}B_{c}^{-}\right] =(17.7\pm
6.7)~\mathrm{MeV}.  \label{eq:DW2}
\end{equation}


\section{Decays into $D$ meson pairs}

\label{sec:Widths2}


It has been noted above that the hadronic molecule $\mathcal{M}$ can decay
to charmed mesons due to annihilation of the $b\overline{b}$ pair in $%
\mathcal{M}$. We study here the channels in which $\mathcal{M}$ transforms
to mesons $D^{(\ast )+}D^{(\ast )-}$, $D^{(\ast )0}\overline{D}^{(\ast )0}$,
and $D_{s}^{+(\ast )}D_{s}^{-(\ast )}$. In the context of the SR method
these processes can be explored by means of the three-point correlation
functions in which the heavy quark condensate $\langle \overline{b}b\rangle $
is replaced by $\langle \alpha _{s}G^{2}/\pi \rangle $.


\subsection{$\mathcal{M}\rightarrow D^{0}\overline{D}^{0}$ and $D^{+}D^{-}$}


Let us analyze in a detailed form the decay $\mathcal{M}\rightarrow D^{0}%
\overline{D}^{0}$. In this process the coupling $\widetilde{g}_{1}$ that
describes strong interaction of particles $\mathcal{M}$, $D^{0}$, and $%
\overline{D}^{0}$ at the vertex $\mathcal{M}D^{0}\overline{D}^{0}$ can be
obtained from the three-point correlation function%
\begin{eqnarray}
\widetilde{\Pi }_{1}(p,p^{\prime }) &=&i^{2}\int d^{4}xd^{4}ye^{ip^{\prime
}y}e^{-ipx}\langle 0|\mathcal{T}\{J^{D^{0}}(y)  \notag \\
&&\times J^{\overline{D}^{0}}(0)J^{\dagger }(x)\}|0\rangle .  \label{eq:CF1A}
\end{eqnarray}%
Above $J^{D^{0}}(x)$ and $J^{\overline{D}^{0}}(x)$ are the interpolating
currents for the mesons $D^{0}$ and $\overline{D}^{0}$
\begin{equation}
J^{D^{0}}(x)=\overline{u}_{j}(x)i\gamma _{5}c_{j}(x),\ J^{\overline{D}%
^{0}}(x)=\overline{c}_{i}(x)i\gamma _{5}u_{i}(x).  \label{eq:CRB}
\end{equation}

\begin{figure}[h]
\includegraphics[width=8.5cm]{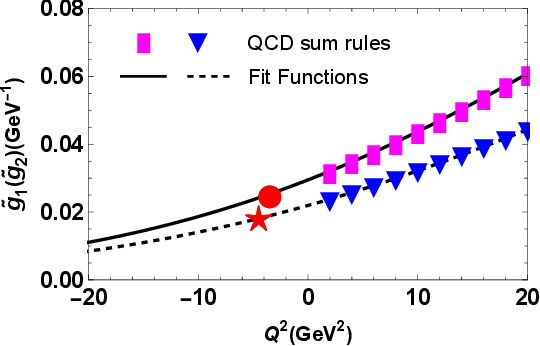}
\caption{Sum rules results and extrapolating functions for the form factors $%
\widetilde{g}_{1}(Q^{2})$ (solid line), $\widetilde{g}_{2}(Q^{2})$ (dashed
line). The circle and star mark the points $Q^{2}=-m_{D^0}^{2}$, and $%
Q^{2}=-m_{D^{\ast 0}}^{2}$, respectively. }
\label{fig:Fit1}
\end{figure}

First, we find the physical side of the SR and write down $\widetilde{\Pi }%
_{1}(p,p^{\prime })$ in the following way
\begin{eqnarray}
&&\widetilde{\Pi }_{1}^{\mathrm{Phys}}(p,p^{\prime })=\frac{\langle
0|J^{D^{0}}|D^{0}(p^{\prime })\rangle }{p^{\prime 2}-m_{D^{0}}^{2}}\frac{%
\langle 0|J^{\overline{D}^{0}}|\overline{D}^{0}(q)\rangle }{%
q^{2}-m_{D^{0}}^{2}}  \notag  \label{eq:CF2} \\
&&\times \langle D^{0}(p^{\prime })\overline{D}^{0}(q)|\mathcal{M}(p)\rangle
\frac{\langle \mathcal{M}(p)|J^{\dagger }|0\rangle }{p^{2}-m^{2}}+\cdots ,
\notag \\
&&
\end{eqnarray}%
where $m_{D^{0}}=$ $(1864.84\pm 0.05)~\mathrm{MeV}$ is the $D^{0}$ and $%
\overline{D}^{0}$ mesons' mass.

To simplify $\widetilde{\Pi }_{1}^{\mathrm{Phys}}(p,p^{\prime })$, we
rewrite it in terms of the involved particles' masses and current couplings
(decay constants). The matrix element of the pseudoscalar $D^{0}$ ($%
\overline{D}^{0}$) meson is given by the expression%
\begin{equation}
\langle 0|J^{D^{0}}|D^{0}\rangle =\frac{f_{D}m_{D^{0}}^{2}}{m_{c}},
\end{equation}%
with $f_{D}=(211.9\pm 1.1)~\mathrm{MeV}$ being its decay constant \cite%
{Rosner:2015wva}. The vertex $\langle D^{0}(p^{\prime })\overline{D}^{0}(q)|%
\mathcal{M}(p)\rangle $ is equal to
\begin{equation}
\langle D^{0}(p^{\prime })\overline{D}^{0}(q)|\mathcal{M}(p)\rangle =%
\widetilde{g}_{1}(q^{2})p\cdot p^{\prime }.  \label{eq:ME3B}
\end{equation}%
Here, $\widetilde{g}_{1}(q^{2})$ is the strong form factor that at the mass
shell of the $\overline{D}^{0}$ meson, i.e., at $q^{2}=m_{D^{0}}^{2}$ gives
the strong coupling $G_{1}$.

Afterwards, we recast $\widetilde{\Pi }_{1}^{\mathrm{Phys}}(p,p^{\prime })$
into the following form
\begin{eqnarray}
&&\widetilde{\Pi }_{1}^{\mathrm{Phys}}(p,p^{\prime })=\widetilde{g}%
_{1}(q^{2})\frac{\Lambda f_{D}^{2}m_{D}^{4}}{2m_{c}^{2}\left(
p^{2}-m^{2}\right) \left( p^{\prime 2}-m_{D^{0}}^{2}\right) }  \notag \\
&&\times \frac{\left( m^{2}+m_{D^{0}}^{2}-q^{2}\right) }{%
(q^{2}-m_{D^{0}}^{2})}+\cdots ,  \label{eq:CorrF5}
\end{eqnarray}%
Note that the whole expression in Eq.\ (\ref{eq:CorrF5}) is the invariant
amplitude $\widetilde{\Pi }_{1}^{\mathrm{Phys}}(p^{2},p^{\prime 2},q^{2})$
which can be employed to get SR for the form factor $\widetilde{g}%
_{1}(q^{2}) $.

The second component necessary to find SR for $\widetilde{g}_{1}(q^{2})$ is
the correlator Eq.\ (\ref{eq:CF1A}) calculated in terms of the quark
propagators, which reads%
\begin{eqnarray}
&&\widetilde{\Pi }_{1}^{\mathrm{OPE}}(p,p^{\prime })=\frac{1}{3}\int
d^{4}xd^{4}ye^{ip^{\prime }y}e^{-ipx}\langle \overline{b}b\rangle  \notag \\
&&\times \mathrm{Tr}\left[ \gamma _{5}S_{c}^{ja}(y-x){}S_{c}^{ai}(x)\gamma
_{5}S_{u}^{ij}(-y)\right] ,  \label{eq:QCDsideA}
\end{eqnarray}%
where $S_{u}(x)$ is the $u$ quark propagator \cite{Agaev:2020zad}.

The correlation function $\widetilde{\Pi }_{1}^{\mathrm{OPE}}(p,p^{\prime })$
depends on three propagators and vacuum condensate $\langle \overline{b}%
b\rangle $ of $b$ quarks. This function differs from  standard correlators
which contain four heavy quark propagators $S_{b(c)}(x)$. In fact, to
calculate $\widetilde{\Pi }_{1}^{\mathrm{OPE}}(p,p^{\prime })$ one contracts
heavy and light quark fields, and because mesons $D^{0}\overline{D}^{0}$ do
not contain $b$ quarks gets an expression with the condensate $\langle
\overline{b}b\rangle $.

For further studies, we make use of the relation between condensates
\begin{equation}
\langle \overline{b}b\rangle =-\frac{1}{12m_{b}}\langle \frac{\alpha
_{s}G^{2}}{\pi }\rangle  \label{eq:Conden}
\end{equation}%
derived in Ref.\ \cite{Shifman:1978bx} from the sum rule analysis. This
expression was obtained at the leading order of the perturbative QCD and is
valid provided that the higher order corrections in $m_{b}^{-1}$ are very small.
Having activated this equality, we get
\begin{eqnarray}
&&\widetilde{\Pi }_{1}^{\mathrm{OPE}}(p,p^{\prime })=-\frac{1}{36m_{b}}%
\langle \frac{\alpha _{s}G^{2}}{\pi }\rangle \int d^{4}xd^{4}ye^{ip^{\prime
}y}e^{-ipx}  \notag \\
&&\times \mathrm{Tr}\left[ \gamma _{5}S_{c}^{ja}(y-x){}S_{c}^{ai}(x)\gamma
_{5}S_{u}^{ij}(-y)\right] .  \label{eq:CF6A}
\end{eqnarray}%
Stated differently, the correlator $\widetilde{\Pi }_{1}^{\mathrm{OPE}%
}(p,p^{\prime })$ is suppressed by the factor $\langle \alpha _{s}G^{2}/\pi
\rangle $. We denote the corresponding invariant amplitude as $\widetilde{%
\Pi }_{1}^{\mathrm{OPE}}(p^{2},p^{\prime 2},q^{2})$ and carry out all
required operations to fix SR for the form factor $\widetilde{g}_{1}(q^{2})$.

We omit standard details and present only final results for the process $%
\mathcal{M}\rightarrow D^{0}\overline{D}^{0}$. Numerical computations of the
form factor $\widetilde{g}_{1}(q^{2})$ are performed using
\begin{equation}
M_{2}^{2}\in \lbrack 1.5,3]~\mathrm{GeV}^{2},\ s_{0}^{\prime }\in \lbrack
5,5.2]~\mathrm{GeV}^{2}.
\end{equation}%
which are the Borel and continuum subtraction parameters in the $D^{0}$
channel. The coupling $\widetilde{g}_{1}$ is fixed at the mass shell $%
q^{2}=m_{D^{0}}^{2}$ and is equal to
\begin{equation}
\widetilde{g}_{1}\equiv \widetilde{\mathcal{G}}_{1}(-m_{D^{0}}^{2})=(2.5\pm
0.4)\times 10^{-2}\ \mathrm{GeV}^{-1}.  \label{eq:G1}
\end{equation}%
To find $\widetilde{g}_{1}$, we have utilized the SR data in the interval $%
Q^{2}=2-20\ \mathrm{GeV}^{2}$ and the fit function with parameters $%
\widetilde{\mathcal{G}}_{1}^{0}=0.03~\mathrm{GeV}^{-1}$, $\widetilde{a}%
_{1}^{1}=6.89$, and $\widetilde{a}_{1}^{2}=-8.36$. The SR data and
extrapolating function $\widetilde{\mathcal{G}}_{1}(Q^{2})$ are plotted in
Fig.\ \ref{fig:Fit1}.

The width of this channel is computed using the formula
\begin{equation}
\Gamma \left[ \mathcal{M}\rightarrow D^{0}\overline{D}^{0}\right] =%
\widetilde{g}_{1}^{2}\frac{m_{D^{0}}^{2}\widetilde{\lambda }_{1}}{8\pi }%
\left( 1+\frac{\widetilde{\lambda }_{1}^{2}}{m_{D^{0}}^{2}}\right) ,
\end{equation}%
where $\widetilde{\lambda }_{1}=\lambda (m,m_{D^{0}}^{2},m_{D^{0}}^{2})$.
Finally, we get
\begin{equation}
\Gamma \left[ \mathcal{M}\rightarrow D^{0}\overline{D}^{0}\right] =(6.1\pm
1.4)~\mathrm{MeV}.
\end{equation}

The strong coupling and width of the process $\mathcal{M}\rightarrow
D^{+}D^{-}$ are close to ones of the decay $\mathcal{M}\rightarrow D^{0}%
\overline{D}^{0}$. Modifications connected with the mass difference of the
mesons $D^{0}$ and $D^{\pm }$ are small and neglected in the present paper:
We set $\Gamma \left[ \mathcal{M}\rightarrow D^{+}D^{-}\right] \approx
\Gamma \left[ \mathcal{M}\rightarrow D^{0}\overline{D}^{0}\right] .$


\subsection{$\mathcal{M}\rightarrow D^{\ast 0}\overline{D}^{\ast 0}$ and $%
D^{\ast +}D^{\ast -}$}


The decay which we are going to consider in this subsection is the process $%
\mathcal{M}\rightarrow D^{\ast 0}\overline{D}^{\ast 0}$. Analysis of this
channel differs from our previous studies only by some technical details. In
the case of the decay $\mathcal{M}\rightarrow D^{\ast 0}\overline{D}^{\ast
0} $ we start from exploration of the correlator
\begin{eqnarray}
&&\widetilde{\Pi }_{2\mu \nu }(p,p^{\prime })=i^{2}\int
d^{4}xd^{4}ye^{ip^{\prime }y}e^{-ipx}\langle 0|\mathcal{T}\{J_{\mu
}^{D^{\ast 0}}(y)  \notag \\
&&\times J_{\nu }^{\overline{D}^{\ast 0}}(0)J^{\dagger }(x)\}|0\rangle .
\end{eqnarray}%
The correlation function $\widetilde{\Pi }_{2\mu \nu }(p,p^{\prime })$
contains the interpolating currents $J_{\mu }^{D^{\ast 0}}(x)$ and $J_{\nu
}^{\overline{D}^{\ast 0}}(x)$ of the vector mesons $D^{\ast 0}$ and $%
\overline{D}^{\ast 0}$, respectively:
\begin{equation}
J_{\mu }^{D^{\ast 0}}(x)=\overline{u}_{i}(x)\gamma _{\mu }c_{i}(x),\ J_{\nu
}^{\overline{D}^{\ast 0}}(x)=\overline{c}_{j}(x)\gamma _{\nu }u_{j}(x).
\end{equation}

In terms of the parameters of the ground-level particles $\widetilde{\Pi }%
_{2\mu \nu }(p,p^{\prime })$ is given by the expression
\begin{eqnarray}
&&\widetilde{\Pi }_{2\mu \nu }^{\mathrm{Phys}}(p,p^{\prime })=\frac{\langle
0|J_{\mu }^{D^{\ast 0}}|D^{\ast 0}(p^{\prime },\varepsilon _{1})\rangle }{%
p^{\prime 2}-m_{D^{\ast 0}}^{2}}\frac{\langle 0|J_{\nu }^{\overline{D}^{\ast
0}}|\overline{D}^{\ast 0}(q,\varepsilon _{2})\rangle }{q^{2}-m_{D^{\ast
0}}^{2}}  \notag \\
&&\times \langle D^{\ast 0}(p^{\prime },\varepsilon _{1})\overline{D}^{\ast
0}(q,\varepsilon _{2})|\mathcal{M}(p)\rangle \frac{\langle \mathcal{M}%
(p)|J^{\dagger }|0\rangle }{p^{2}-m^{2}}+\cdots ,  \notag \\
&&
\end{eqnarray}%
where $m_{D^{\ast 0}}=(2006.85\pm 0.05)\ \mathrm{MeV}$ is the mass of the
mesons $D^{\ast 0}$ and $\overline{D}^{\ast 0}$. To continue let us utilize
the matrix elements
\begin{eqnarray}
&&\langle 0|J_{\mu }^{D^{\ast 0}}|D^{\ast 0}(p^{\prime },\varepsilon
_{1})\rangle =f_{D^{\ast }}m_{D^{\ast 0}}\varepsilon _{1\mu },  \notag \\
&&\langle 0|J_{\nu }^{\overline{D}^{\ast 0}}|\overline{D}^{\ast
0}(q,\varepsilon _{2})\rangle =f_{D^{\ast }}m_{D^{\ast 0}}\varepsilon _{2\nu
},
\end{eqnarray}%
with $f_{D^{\ast }}=(00\pm 0.05)\ \mathrm{MeV}$ and $\varepsilon _{1}$, $%
\varepsilon _{2}$ being the decay constant and polarization vectors of the
mesons $D^{\ast 0}$ and $\overline{D}^{\ast 0}$. The vertex $\mathcal{M}%
D^{\ast 0}\overline{D}^{\ast 0}$ has the following form%
\begin{eqnarray}
&&\langle D^{\ast 0}(p^{\prime },\varepsilon _{1})\overline{D}^{\ast
0}(q,\varepsilon _{2})|\mathcal{M}(p)\rangle =\widetilde{g}_{2}(q^{2})\left[
q\cdot p^{\prime }\right.  \notag \\
&&\left. \times \varepsilon _{1}^{\ast }\cdot \varepsilon _{2}^{\ast
}-q\cdot \varepsilon _{1}^{\ast }p^{\prime }\cdot \varepsilon _{2}^{\ast }
\right] .
\end{eqnarray}%
Then $\widetilde{\Pi }_{2\mu \nu }(p,p^{\prime })$ expressed using
parameters of the hadronic molecule $\mathcal{M}$, and mesons $D^{\ast 0}$,
and $\overline{D}^{\ast 0}$ is given by the formula
\begin{eqnarray}
&&\widetilde{\Pi }_{2\mu \nu }^{\mathrm{Phys}}(p,p^{\prime })=\frac{%
\widetilde{g}_{2}(q^{2})\Lambda f_{D^{\ast }}^{2}m_{D^{\ast 0}}^{2}}{\left(
p^{2}-m^{2}\right) \left( p^{\prime 2}-m_{D^{\ast 0}}^{2}\right)
(q^{2}-m_{D^{\ast 0}}^{2})}  \notag \\
&&\times \left[ \frac{1}{2}\left( m^{2}-m_{D^{\ast 0}}^{2}-q^{2}\right)
g_{\mu \nu }-q_{\mu }p_{\nu }^{\prime }\right] +\cdots .
\end{eqnarray}%
The correlator $\widetilde{\Pi }_{2\mu \nu }(p,p^{\prime })$ computed by
employing the quark propagators reads%
\begin{eqnarray}
&&\widetilde{\Pi }_{2\mu \nu }^{\mathrm{OPE}}(p,p^{\prime })=i^{2}\int
d^{4}xd^{4}ye^{ip^{\prime }y}e^{-ipx}\langle \overline{b}b\rangle  \notag \\
&&\times \mathrm{Tr}\left[ \gamma _{\mu
}S_{c}^{ja}(y-x){}S_{c}^{ai}(x)\gamma _{\nu }S_{u}^{ij}(-y)\right] .
\end{eqnarray}%
The correlator $\widetilde{\Pi }_{2\mu \nu }(p,p^{\prime })$ in both cases
contains the two Lorentz structures $g_{\mu \nu }$ and $q_{\mu }p_{\nu
}^{\prime }$. We use invariant amplitudes that correspond to structures
proportional to $g_{\mu \nu }$ and derive a required sum rule.

Numerical analysis has been carried out using the following input
parameters: For the Borel and continuum subtraction parameters in the
channel of the meson $D^{\ast 0}$ we use
\begin{equation}
M_{2}^{2}\in \lbrack 2,3]~\mathrm{GeV}^{2},\ s_{0}^{\prime }\in \lbrack
5.7,5.8]~\mathrm{GeV}^{2}.
\end{equation}%
The decay constant $f_{D^{\ast }}$ of the mesons $D^{\ast 0}$ is fixed as $%
(252.2\pm 22.66)~\mathrm{MeV}$ . The extrapolating function $\widetilde{%
\mathcal{G}}_{2}(Q^{2})$ has the parameters: $\widetilde{\mathcal{G}}%
_{2}^{0}=0.022~\mathrm{GeV}^{-1}$, $\widetilde{a}_{2}^{1}=6.70$, and $%
\widetilde{a}_{2}^{2}=-8.18$\ (see, Fig.\ \ref{fig:Fit1} ). Then, the strong
coupling $\widetilde{g}_{2}$ amounts to
\begin{equation}
\widetilde{g}_{2}\equiv \widetilde{\mathcal{G}}_{2}(-m_{D^{\ast
0}}^{2})=(1.8\pm 0.4)\times 10^{-2}\ \mathrm{GeV}^{-1}.
\end{equation}%
The partial width of the decay $\mathcal{M}\rightarrow D^{\ast 0}\overline{D}%
^{\ast 0}$ can be found by means of the formula
\begin{eqnarray}
&&\Gamma \left[ \mathcal{M}\rightarrow D^{\ast 0}\overline{D}^{\ast 0}\right]
=\frac{\widetilde{g}_{2}^{2}\widetilde{\lambda }_{2}}{16\pi }\left(
m^{2}-4m_{D^{\ast 0}}^{2}+6\frac{m_{D^{\ast 0}}^{4}}{m^{2}}\right) ,  \notag
\\
&&
\end{eqnarray}%
where $\lambda _{3}=\lambda (m,m_{D^{\ast 0}},m_{D^{\ast 0}})$. The partial
decay width of this decay is

\begin{equation}
\Gamma \left[ \mathcal{M}\rightarrow D^{\ast 0}\overline{D}^{\ast 0}\right]
=(5.7\pm 1.8)~\mathrm{MeV}.
\end{equation}%
The width of the process $\mathcal{M}\rightarrow $ $D^{\ast +}D^{\ast -}$
equals approximately to $\Gamma \left[ \mathcal{M}\rightarrow D^{\ast 0}%
\overline{D}^{\ast 0}\right] $.

\bigskip


\subsection{$\mathcal{M}\rightarrow D_{s}^{(\ast )+}D_{s}^{(\ast )-}$}


The analysis of the decay $\mathcal{M}\rightarrow D_{s}^{+}D_{s}^{-}$ is
done using the scheme which has been explained and applied above. There are
modifications due to the parameters of the $D_{s}^{\pm }$ mesons such as
their masses $m_{D_{s}}=(1969.0\pm 1.4)~\mathrm{MeV}$ and decay constant $%
f_{D_{s}}=(249.9\pm 0.5)~\mathrm{MeV}$. The SR data are calculated by
employing the Borel and continuum subtraction parameters in the $D_{s}^{+}$
channel
\begin{equation}
M_{2}^{2}\in \lbrack 2.5,3.5]~\mathrm{GeV}^{2},\ s_{0}^{\prime }\in \lbrack
6,8]~\mathrm{GeV}^{2}.
\end{equation}%
The extrapolating function $\widetilde{\mathcal{G}}_{3}(Q^{2})$ is
determined by $\widetilde{\mathcal{G}}_{3}^{0}=0.024~\mathrm{GeV}^{-1}$, $%
\widetilde{a}_{3}^{1}=6.88$, and $\widetilde{a}_{3}^{2}=-8.42$. The strong
coupling $\widetilde{g}_{3}$ is found by means of this function and is equal
to \
\begin{equation}
\widetilde{g}_{3}\equiv \widetilde{\mathcal{G}}_{3}(-m_{D_{s}}^{2})=(2.1\pm
0.4)\times 10^{-2}\ \mathrm{GeV}^{-1}.
\end{equation}%
The width of the decay $\mathcal{M}\rightarrow D_{s}^{+}D_{s}^{-}$ amounts
to
\begin{equation}
\Gamma \left[ \mathcal{M}\rightarrow D_{s}^{+}D_{s}^{-}\right] =(4.1\pm 1.2)~%
\mathrm{MeV}.
\end{equation}

In the case of the second decay $\mathcal{M}\rightarrow D_{s}^{\ast
+}D_{s}^{\ast -}$ the corresponding strong coupling $\widetilde{g}_{4}$ is%
\begin{equation}
\widetilde{g}_{4}\equiv \widetilde{\mathcal{G}}_{4}(-m_{D_{s}^{\ast
}}^{2})=(1.6\pm 0.3)\times 10^{-2}\ \mathrm{GeV}^{-1}.
\end{equation}%
Here, the parameters of the function $\widetilde{\mathcal{G}}_{4}(Q^{2})$
are $\widetilde{\mathcal{G}}_{4}^{0}=0.019~\mathrm{GeV}^{-1}$, $\widetilde{a}%
_{4}^{1}=6.703$, and $\widetilde{a}_{4}^{2}=-8.18$. \ For the width of this
process we find

\begin{equation}
\Gamma \left[ \mathcal{M}\rightarrow D_{s}^{\ast +}D_{s}^{\ast -}\right]
=(4.4\pm 1.2)~\mathrm{MeV}.
\end{equation}%
In calculating of the coupling $\widetilde{g}_{4}$ and width of the decay $%
\mathcal{M}\rightarrow D_{s}^{\ast +}D_{s}^{\ast -}$ as the mass and the
decay constant of the mesons $D_{s}^{\ast \pm }$ we have utilized $%
m_{D_{s}^{\ast }}=(2112.2\pm 0.4)~\mathrm{MeV}$ and $f_{D_{s}^{\ast
}}=(221\pm 7)~\mathrm{MeV}$, respectively.

\begin{quote}
\end{quote}

\section{Decays into $B^{(\ast )+}B^{(\ast )-}$, $B^{(\ast )0}\overline{B}%
^{(\ast )0}$, and $B_{s}^{(\ast )0}\overline{B}_{s}^{(\ast )0}$ mesons}

\label{sec:Widths3}


Decays of the hadronic molecule $\mathcal{M}$ into pairs of the $B$ and $%
B_{s}$ mesons can be studied by the same way. These processes are triggered
by $c\overline{c}$ annihilation in the molecule and lead to creations of $BB$
meson pairs with appropriate charges and quantum numbers. One can see that $%
\mathcal{M}\rightarrow B^{(\ast )+}B^{(\ast )-}$, $B^{(\ast )0}\overline{B}%
^{(\ast )0}$, and $B_{s}^{0}\overline{B}_{s}^{0}$ are possible decay
channels of the hadronic molecule $\mathcal{M}$.

The correlation function required to estimate the strong coupling $G_{1}\ $%
at the vertex $\mathcal{M}B^{+}B^{-}$ is given by the formula

\begin{eqnarray}
&&\widehat{\Pi }_{1}(p,p^{\prime })=i^{2}\int d^{4}xd^{4}ye^{ip^{\prime
}y}e^{-ipx}\langle 0|\mathcal{T}\{J^{B^{-}}(y)  \notag \\
&&\times J^{B^{+}}(0)J^{\dagger }(x)\}|0\rangle ,
\end{eqnarray}%
where $J^{B^{+}}(x)$ and $J^{B^{-}}(x)$ are the interpolating currents of
the mesons $B^{+}$ and $B^{-}$
\begin{equation}
J^{B^{+}}(x)=\overline{b}_{i}(x)i\gamma _{5}u_{i}(x),\ J^{B^{-}}(x)=%
\overline{u}_{j}(x)i\gamma _{5}b_{j}(x).
\end{equation}%
The contribution of the ground-state particles $\mathcal{M}$, $B^{+}$, and $%
B^{-}$ to $\widehat{\Pi }_{1}(p,p^{\prime })$ is
\begin{eqnarray}
&&\widehat{\Pi }_{1}^{\mathrm{Phys}}(p,p^{\prime })=\frac{\langle
0|J^{B^{-}}|B^{-}(p^{\prime })\rangle }{p^{\prime 2}-m_{B}^{2}}\frac{\langle
0|J^{B^{+}}|B^{+}(q)\rangle }{q^{2}-m_{B}^{2}}  \notag \\
&&\times \langle B^{-}(p^{\prime })B^{+}(q)|\mathcal{M}(p)\rangle \frac{%
\langle \mathcal{M}(p)|J^{\dagger }|0\rangle }{p^{2}-m^{2}}+\cdots .  \notag
\\
&&
\end{eqnarray}%
To compute the phenomenological side of the SR $\widehat{\Pi }_{1}^{\mathrm{%
Phys}}(p,p^{\prime })$, we make use of the matrix elements
\begin{equation}
\langle 0|J^{B^{\pm }}|B^{\pm }\rangle =\frac{f_{B}m_{B}^{2}}{m_{b}},
\label{eq:ME2B}
\end{equation}%
and
\begin{equation}
\langle B^{-}(p^{\prime })B^{+}(q)|\mathcal{M}\rangle =G_{1}(q^{2})p\cdot
p^{\prime }.  \label{eq:ME3C}
\end{equation}%
In the formulas above, $m_{B}=(2112.2\pm 0.4)~\mathrm{MeV}$ and $%
f_{B}=(206\pm 7)~\mathrm{MeV}$ are the mass and decay constant of the mesons
$B^{\pm }$.

Then for $\widehat{\Pi }_{1}^{\mathrm{Phys}}(p,p^{\prime })$, we get
\begin{eqnarray}
&&\widehat{\Pi }_{1}^{\mathrm{Phys}}(p,p^{\prime })=\frac{%
G_{1}(q^{2})\Lambda f_{B}^{2}m_{B}^{4}}{m_{b}^{2}\left( p^{2}-m^{2}\right)
\left( p^{\prime 2}-m_{B}^{2}\right) }  \notag \\
&&\times \frac{m^{2}+m_{B}^{2}-q^{2}}{2(q^{2}-m_{B}^{2})}+\cdots .
\end{eqnarray}%
The correlation function $\widehat{\Pi }_{1}(p,p^{\prime })$ in terms of the
quark propagators reads
\begin{eqnarray}
&&\widehat{\Pi }_{1}^{\mathrm{OPE}}(p,p^{\prime })=\frac{1}{3}\int
d^{4}xd^{4}ye^{ip^{\prime }y}e^{-ipx}\langle \overline{c}c\rangle  \notag \\
&&\times \mathrm{Tr}\left[ \gamma _{5}S_{b}^{ja}(y-x){}S_{b}^{ai}(x)\gamma
_{5}S_{u}^{ij}(-y)\right] .
\end{eqnarray}%
The heavy quark condensate $\langle \overline{c}c\rangle $ with some
accuracy can be replaced by \cite{Shifman:1978bx}
\begin{equation}
\langle \overline{c}c\rangle =-\frac{1}{12m_{c}}\langle \frac{\alpha
_{s}G^{2}}{\pi }\rangle .
\end{equation}%
This leads to the correlator
\begin{eqnarray}
&&\widehat{\Pi }_{1}^{\mathrm{OPE}}(p,p^{\prime })=-\frac{1}{36m_{c}}\langle
\frac{\alpha _{s}G^{2}}{\pi }\rangle \int d^{4}xd^{4}ye^{ip^{\prime
}y}e^{-ipx}  \notag \\
&&\times \mathrm{Tr}\left[ \gamma _{5}S_{b}^{ja}(y-x){}S_{b}^{ai}(x)\gamma
_{5}S_{u}^{ij}(-y)\right] .
\end{eqnarray}

Numerical computations of the form factor $G_{1}(q^{2})$ are performed for $%
- $ $q^{2}=Q^{2}=2-10~\mathrm{GeV}^{2}$. The coupling $G_{1}$ is extracted
at the $Q^{2}=-m_{B}^{2}$ by utilizing the fit function $\widehat{\mathcal{G}%
}_{1}(Q^{2})$ with parameters $\widehat{\mathcal{G}}_{1}^{0}=0.045~\mathrm{%
GeV}^{-1}$, $\widehat{a}_{2}^{1}=2.82$, and $\widehat{a}_{4}^{2}=-2.61$
\begin{equation}
G_{1}\equiv \widehat{\mathcal{G}}_{1}(-m_{B}^{2})=(2.6\pm 0.4)\times
10^{-2}\ \mathrm{GeV}^{-1}.
\end{equation}%
Then the width of the process $\mathcal{M\rightarrow }B^{+}B^{-}$ is equal to

\begin{equation}
\Gamma \left[ \mathcal{M\rightarrow }B^{+}B^{-}\right] =(3.8\pm 0.9)~\mathrm{%
MeV}.
\end{equation}

In Table we provide brief information about the remaining decay modes of the
molecule $\mathcal{M}$.

\begin{table}[tbp]
\begin{tabular}{|c|c|c|c|}
\hline\hline
$i$ & Channels & ${G}_{i}\times10^2~(\mathrm{GeV}^{-1})$ & $\Gamma_{i}~(%
\mathrm{MeV})$ \\ \hline
$1$ & $B^{+}B^{-}$ & $2.6 \pm 0.4$ & $3.8 \pm 0.9$ \\
$2$ & $B^{\ast +}B^{\ast -}$ & $2.8 \pm 0.3$ & $4.2 \pm 1.3$ \\
$3$ & $B_{s}^{0}\overline{B}_{s}^{0}$ & $2.0 \pm 0.3$ & $2.2 \pm 0.5 $ \\
$4$ & $B_{s}^{\ast 0}\overline{B}_{s}^{\ast 0}$ & $2.6 \pm 0.5$ & $3.5 \pm
1.0$ \\ \hline\hline
\end{tabular}%
\caption{Decay modes of the hadronic molecule $\mathcal{M}$ generated by $c%
\overline{c}$ annihilation, strong couplings $G_{i}$, and partial widths $%
\Gamma _{i}$.}
\label{tab:Channels}
\end{table}
Information on the partial widths of the channels considered in two last
sections permit us to evaluate the full width of the hadronic molecule
\begin{equation}
\Gamma \left[ \mathcal{M}\right] =(155\pm 23)~\mathrm{MeV}.
\end{equation}


\section{Concluding remarks}

\label{sec:Conc}


In this article, we have studied the scalar hadronic molecule $\mathcal{M}%
=B_{c}^{+}B_{c}^{-}$ in the context of the QCD sum rule method. Our prediction
for the mass of this molecule $m=(12725\pm 85)~\mathrm{MeV}$ provides
important information about this state. It turns out that the molecular
structure $\mathcal{M}$ is unstable against two-meson fall-apart processes.

The fully heavy hadronic molecules $B_{c}^{(\ast )+}B_{c}^{(\ast )-}$ were
explored in the framework of the coupled-channel unitary approach in Ref.\
\cite{Liu:2023gla}. Results in this method depend on the parameter $\Lambda $
used to regularize some integrals. In the case of the scalar structure $%
B_{c}^{+}B_{c}^{-}$ at $\Lambda =600~\mathrm{MeV}$,  the authors found the
pole position $E_{\mathrm{p}}=(12503.3-i126.8)~\mathrm{MeV}$. In other
words, the mass of the hadronic molecule $B_{c}^{+}B_{c}^{-}$ is $12503.3~%
\mathrm{MeV}$,  which is smaller than the one obtained in the present work.
Nevertheless, \ the molecule $B_{c}^{+}B_{c}^{-}$ in this picture is also
unstable and can  decay strongly into the pair of mesons $\eta _{b}\eta _{c}$.

The mass of $\mathcal{M}$ has enabled us to fix its possible decay channels.
We have computed partial widths of the two-meson dissociations $\mathcal{M}%
\rightarrow \eta _{b}\eta _{c},\ J/\psi \Upsilon $, and $B_{c}^{+}B_{c}^{-}$%
, which are dominant decay modes of the molecule $\mathcal{M}$. The
annihilations of the $b\overline{b}$ and $c\overline{c}$ pairs in $\mathcal{M%
}$ give rise to other two-meson strong decays: We have considered numerous
processes generated by this second mechanism. To calculate the widths of
two-meson decays, we have employed the method of the three-point sum rule.  This is necessary to estimate strong couplings of particles at
various $\mathcal{M}$-meson-meson vertices. The full width of $\mathcal{M}$
saturated by these decay modes amounts to $\Gamma \left[ \mathcal{M}\right]
=(155\pm 23)~\mathrm{MeV}$: The fall-apart processes constitute $66\%$ of
this parameter.

In our previous publications \cite{Agaev:2024wvp,Agaev:2024mng,Agaev:2025},
we performed comprehensive investigations of the four-quark mesons $0^{+}$, $%
1^{+}$ and $2^{+}$ with the quark content $bc\overline{b}\overline{c}$. \
These states were treated as diquark-antidiquark structures. The scalar
particle $T_{\mathrm{bc\overline{b}\overline{c}}}$ was modeled as a
tetraquark composed of the scalar diquark and antidiquark \cite%
{Agaev:2024wvp}. The mass and full width of this tetraquark were found equal
to $m=(12697\pm 90)~\mathrm{MeV}$ and $\Gamma \left[ T_{\mathrm{bc\overline{b%
}\overline{c}}}\right] =(142.4\pm 16.9)~\mathrm{MeV}$, respectively.

Comparing the masses of these two exotic structures, we see that the
hadronic molecule $\mathcal{M}$ is heavier than the diquark-antidiquark
state, though differences between their masses and widths are small: In fact,
within the errors of computations, one may state that these two scalar compounds
have approximately the same masses. Their full widths are also comparable with each
other, though there are identical decay channels with considerably different partial widths.  For instance, the partial width of the decays into $B_{c}^{+}B_{c}^{-}$ mesons  in the  case of the diquark-antidiquark and molecule  structures are  equal to $33.9~
\mathrm{MeV}$ and $17.7~\mathrm{MeV}$,  respectively. Even lower and upper limits for these parameters, obtained  by taking into account the errors of calculations, generate only negligible small  overlapping regions. Therefore,  such decay modes may be used to distinguish the molecular structure of the exotic state  $bc\overline{b}\overline{c}$ from  its diquark organization.

Detailed studies of the diquark-antidiquark and hadronic molecule structures
$bc\overline{b}\overline{c}$ with different spin-parities are necessary to
calculate their parameters and reveal reactions where they may be observed.


\begin{thebibliography}{999}

\bibitem{Jaffe:1976ig} R.~L.~Jaffe,
Phys.\ Rev.\ D \textbf{15}, 267 (1977). 


\bibitem{Bander:1975fb} M.~Bander, G.~L.~Shaw, P.~Thomas, and S.~Meshkov,
Phys.\ Rev.\ Lett.\ \textbf{36}, 695 (1976).


\bibitem{Voloshin:1976ap} M.~B.~Voloshin and L.~B.~Okun,
JETP\ Lett.\ \textbf{23}, 333 (1976).


\bibitem{DeRujula:1976zlg} A.~De~Rujula, H.~Georgi, and S.~L.~Glashow,
Phys.\ Rev.\ Lett.\ \textbf{38}, 317 (1977).


\bibitem{Tornqvist:1991lks} N.~A.~Tornqvist,
Phys.\ Rev.\ Lett.\ \textbf{67}, 556 (1991).


\bibitem{Ding:2008mp} G.~J.~Ding, W.~Huang, J.~F.~Liu, and M.~L.~Yan,
Phys.\ Rev.\ D \textbf{79}, 034026 (2009). 


\bibitem{Zhang:2009vs} J.~R.~Zhang and M.~Q.~Huang,
Phys.\ Rev.\ D \textbf{80}, 056004 (2009). 


\bibitem{Albuquerque:2012rq} R.~M.~Albuquerque, X.~Liu, and M.~Nielsen,
Phys.\ Lett.\ B \textbf{718}, 492 (2012).


\bibitem{Chen:2015ata} W.~Chen, T.~G.~Steele, H.~X.~Chen, and S.~L.~Zhu,
Phys.\ Rev.\ D \textbf{92}, 054002 (2015). 


\bibitem{Karliner:2015ina} M.~Karliner and J.~L.~Rosner,
Phys.\ Rev.\ Lett.\ \textbf{115}, 122001 (2015).


\bibitem{Liu:2016kqx} Y.~Liu and I.~Zahed,
Phys.\ Lett.\ B \textbf{762}, 362 (2016).


\bibitem{Chen:2017vai} R.~Chen, A.~Hosaka, and X.~Liu,
Phys.\ Rev.\ D \textbf{96}, 116012 (2017). 


\bibitem{Sun:2018zqs} Z.~F.~Sun, J.~J.~Xie, and E.~Oset,
Phys.\ Rev.\ D \textbf{97}, 094031 (2018). 


\bibitem{PavonValderrama:2019ixb} M.~Pavon Valderrama,
Eur.\ Phys.\ J.\ A \textbf{56}, 109 (2020).


\bibitem{Molina:2020hde} R.~Molina and E.~Oset,
Phys.\ Lett.\ D \textbf{811}, 135878 (2020) [Erratum: Phys.\ Lett.\ D
\textbf{837}, 137645 (2023)]. 


\bibitem{Xu:2020evn} Y.~J.~Xu, Y.~L.~Liu, C.~Y.~Cui, and M.~Q.~Huang,
Phys.\ Rev.\ D \textbf{104}, 094028 (2021).


\bibitem{Xin:2021wcr} Q.~Xin and Z.~G.~Wang,
Eur.\ Phys.\ J.\ A \textbf{58}, 110 (2022).


\bibitem{Agaev:2022duz} S.~S.~Agaev, K.~Azizi, and H.~Sundu,
J.\ Phys.\ G \textbf{50}, 055002 (2023). 


\bibitem{Agaev:2023eyk} S.~S.~Agaev, K.~Azizi, and H.~Sundu,
Phys.\ Rev.\ D \textbf{107}, 094019 (2023). 


\bibitem{Wang:2023bek} F.~L.~Wang, S.~Q.~Luo, and X.~Liu,
Phys.\ Rev.\ D \textbf{107}, 114017 (2023).


\bibitem{Braaten:2023vgs} E.~Braaten, L.~P.~He, K.~Ingles, and J.~Jiang,
JHEP \textbf{02}, 163 (2024). 


\bibitem{Wu:2023rrp} Q.~Wu, M.~Z.~Liu and L.~S.~Geng,
Eur.\ Phys.\ J.\ C \textbf{84}, 147 (2024).


\bibitem{Liang:2023jxh} W.~H.~Liang, T.~Ban, and E.~Oset,
Phys.\ Rev.\ D \textbf{109}, 054030 (2024). 


\bibitem{Liu:2023gla} W.~Y.~Liu and H.~X.~Chen,
Eur.\ Phys.\ J.\ C \textbf{85}, 636 (2025). 


\bibitem{Liu:2024pio} W.~Y.~Liu and H.~X.~Chen,
Universe \textbf{11}, 36 (2025). 


\bibitem{Wang:2025zss} F.~Wang, G.~Li, S.~D.~Liu, and Q.~Wu,
Phys.\ Rev.\ D \textbf{111}, 094001 (2025).


\bibitem{Braaten:2024tbm} E.~Braaten and R.~Bruschini,
Phys.\ Lett.\ B \textbf{863}, 139386 (2025).


\bibitem{Yalikun:2025ssz} N.~Yalikun, X.~K.~Dong, and U.~G.~Mei\ss {}ner,
Phys. Rev. D \textbf{111}, 094036 (2025)..


\bibitem{LHCb:2020bwg} R.~Aaij \textit{et al.} (LHCb Collaboration),
Sci.\ Bull. \textbf{65}, 1983 (2020).


\bibitem{Bouhova-Thacker:2022vnt} E.~Bouhova-Thacker (ATLAS Collaboration),
PoS \textbf{ICHEP2022}, 806 (2022).


\bibitem{CMS:2023owd} A.~Hayrapetyan, \textit{et al.} (CMS Collaboration)
Phys.\ Rev.\ Lett.\ \textbf{132}, 111901 (2024).


\bibitem{Agaev:2023wua} S.~S.~Agaev, K.~Azizi, B.~Barsbay, and H.~Sundu,
Phys.\ Lett.\ B \textbf{844}, 138089 (2023). 


\bibitem{Agaev:2023ruu} S.~S.~Agaev, K.~Azizi, B.~Barsbay and H.~Sundu,
Eur.\ Phys.\ J. Plus \textbf{138}, 935 (2023). 


\bibitem{Agaev:2023gaq} S.~S.~Agaev, K.~Azizi, B.~Barsbay and H.~Sundu,
Nucl.\ Phys.\ A \textbf{1041}, 122768 (2024). 


\bibitem{Agaev:2023rpj} S.~S.~Agaev, K.~Azizi, B.~Barsbay and H.~Sundu,
Eur.\ Phys.\ J. C \textbf{83}, 994 (2023). 


\bibitem{Aaij:2021vvq} R.~Aaij \textit{et al.} [LHCb Collaboration],
Nature\ Phys.\ \textbf{18}, 751 (2022). 


\bibitem{LHCb:2021auc} R.~Aaij \textit{et al.} [LHCb Collaboration],
Nature\ Commun.\ \textbf{13}, 3351 (2022). 


\bibitem{Agaev:2021vur} S.~S.~Agaev, K.~Azizi, and H.~Sundu,
Nucl.\ Phys.\ B \textbf{975}, 115650 (2022). 


\bibitem{Agaev:2022ast} S.~S.~Agaev, K.~Azizi, and H.~Sundu,
JHEP \textbf{06}, 057 (2022). 


\bibitem{Agaev:2024wvp} S.~S.~Agaev, K.~Azizi, and H.~Sundu,
Phys.\ Lett.\ B \textbf{858}, 139042 (2024). 


\bibitem{Agaev:2024mng} S.~S.~Agaev, K.~Azizi, and H.~Sundu,
Phys.\ Lett.\ B \textbf{864}, 139404 (2025). 


\bibitem{Agaev:2025} S.~S.~Agaev, K.~Azizi, and H.~Sundu,
Phys.\ Rev.\ D \textbf{111}, 074025 (2025). 


\bibitem{Shifman:1978bx} M.~A.~Shifman, A.~I.~Vainshtein and V.~I.~Zakharov,
Nucl.\ Phys.\ B \textbf{147}, 385 (1979).


\bibitem{Shifman:1978by} M.~A.~Shifman, A.~I.~Vainshtein and V.~I.~Zakharov,
Nucl.\ Phys.\ B \textbf{147}, 448 (1979).


\bibitem{Becchi:2020mjz} C.~Becchi, A.~Giachino, L.~Maiani, and
E.~Santopinto,
Phys.\ Lett.\ B \textbf{806}, 135495 (2020).


\bibitem{Becchi:2020uvq} C.~Becchi, A.~Giachino, L.~Maiani, and
E.~Santopinto, 
Phys.\ Lett.\ B \textbf{811}, 135952 (2020). 


\bibitem{Agaev:2023ara} S.~S.~Agaev, K.~Azizi, B.~Barsbay, and H.~Sundu,
Phys.\ Rev.\ D \textbf{109}, 014006 (2024). 


\bibitem{Agaev:2020zad} S.~S.~Agaev, K.~Azizi and H.~Sundu,
Turk.\ J.\ Phys.\ \textbf{44}, 95 (2020). 


\bibitem{PDG:2024} S.~Navas \textit{et al.} [Particle Data Group], Phys.\
Rev.\ D \textbf{110}, 030001 (2024). 


\bibitem{Lakhina:2006vg} O.~Lakhina, and E.~S.~Swanson,
Phys.\ Rev.\ D \textbf{74}, 014012 (2006). 


\bibitem{Wang:2024fwc} Z.~G.~Wang,
Chin.\ Phys.\ C \textbf{48}, 103104 (2024). 


\bibitem{Rosner:2015wva} J.~L.~Rosner, S.~Stone, and R.~S.~Van de
Water,(2015) 
arXiv:1509.02220.
\end{thebibliography}
\end{document}